\documentclass{PoS}

\usepackage{xspace}
\newcommand{\GeVc}{\ensuremath{\mbox{Ge\kern-0.1em V}\!/\!c}\xspace}
\newcommand{\NASixtyOne}{NA61\slash SHINE\xspace}
\newcommand{\dedx}{\ensuremath{{\rm d}E\!/\!{\rm d}x}\xspace}
\newcommand{\Epos}{{\scshape Epos}\xspace}
\newcommand{\Geant}{{\scshape Geant}\xspace}

\title{Two particle correlations from the energy scan with p+p interactions}

\ShortTitle{Two particle correlations from the energy scan with p+p interactions}

\author{\speaker{Bartosz Maksiak} for the \NASixtyOne collaboration%
         \thanks{A footnote may follow.}\\
        Warsaw University of Technology\\
        E-mail: \email{maksiak@if.pw.edu.pl}}

\abstract{
The \NASixtyOne experiment aims to discover the critical point of strongly interacting matter and study the properties of the onset of deconfinement. These goals are to be achieved by performing a two dimensional phase diagram ($T$-$\mu_B$) scan by measurements of hadron production properties in proton-proton, proton-nucleus and nucleus-nucleus interactions as a function of collision energy and system size. Close to the phase transition and/or close to the critical point large fluctuations are predicted. In this contribution preliminary results on two-particle correlations in pseudorapidity and azimuthal angle will be presented for p+p interactions at beam momenta: 20, 31, 40, 80 and 158~\GeVc. The \NASixtyOne results will be compared with the corresponding data of other experiments and model predictions. A striking evolution with collision energy is observed.
}

\FullConference{9th International Workshop on Critical Point and Onset of Deconfinement - CPOD2014,\\
		17-21 November 2014\\
		ZiF (Center of Interdisciplinary Research), University of Bielefeld, Germany}

\begin{document}

\section{Introduction}

Two-particle correlations in $\Delta\eta$, $\Delta\phi$ were studied extensively at RHIC and LHC. They allow to disentangle different sources of correlations: jets, flow, resonance decays, quantum statistics effects, conservation laws, etc. Here we report preliminary results on two-particle $\Delta\eta\Delta\phi$ correlations in inelastic p+p interactions at SPS beam momenta (20, 31, 40, 80, and 158~\GeVc).

\section{$\Delta\eta\Delta\phi$ correlations}

Correlations are calculated as a function of the difference in pseudo-rapidity
($\eta$) and azimuthal angle ($\phi$) between two particles in the same event.
\begin{center}
  \begin{math}
    \Delta\eta = |{\eta}_1 - {\eta}_2|
    \hspace{2cm}
    \Delta\phi = |{\phi}_1 - {\phi}_2|
  \end{math}
\end{center}

The uncorrected ($raw$) correlation function is calculated as:
\begin{equation}
  \label{eq:correlations}
  C^{raw}(\Delta\eta,\Delta\phi)=
  \frac{N_{mixed}^{pairs}}{N_{data}^{pairs}}
  \frac{S(\Delta\eta,\Delta\phi)}{M(\Delta\eta,\Delta\phi)}
\end{equation}
where
\begin{center}
  $S(\Delta\eta,\Delta\phi)=\frac{d^2N^{signal}}{d \Delta \eta d
    \Delta \phi}$; \hspace{0.1cm}
  $M(\Delta\eta,\Delta\phi)=\frac{d^2N^{mixed}}{d \Delta \eta d \Delta
    \phi}$
\end{center}
are the distributions for pairs from data and mixed events, respectively.
The $\Delta\phi$ range is folded, i.e. for $\Delta\phi$ larger than $\pi$ its value is recalculated as $2\pi - \Delta\phi$. In order to allow for a comparison with the RHIC and LHC results the pseudo-rapidity was calculated in the centre-of-mass (CMS) system. The transformation from the laboratory system to the CMS was performed assuming the pion mass for all produced particles. Electrons and positrons were removed by a cut on \dedx, the energy loss of the particle tracks in the TPC detectors.

\begin{figure}[h]
  \centering
  \includegraphics[width=0.25\textwidth]{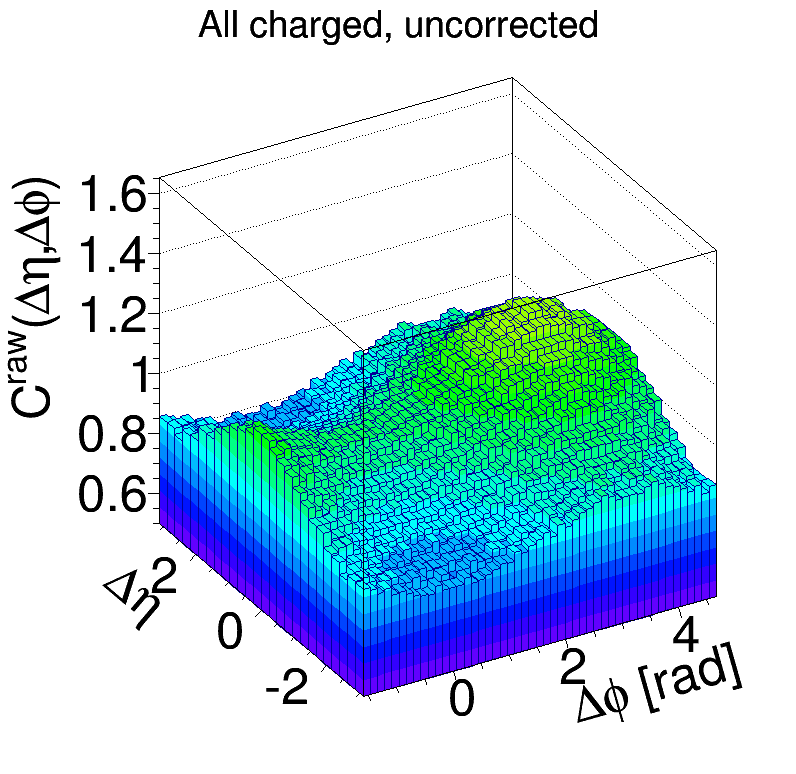}
  \includegraphics[width=0.25\textwidth]{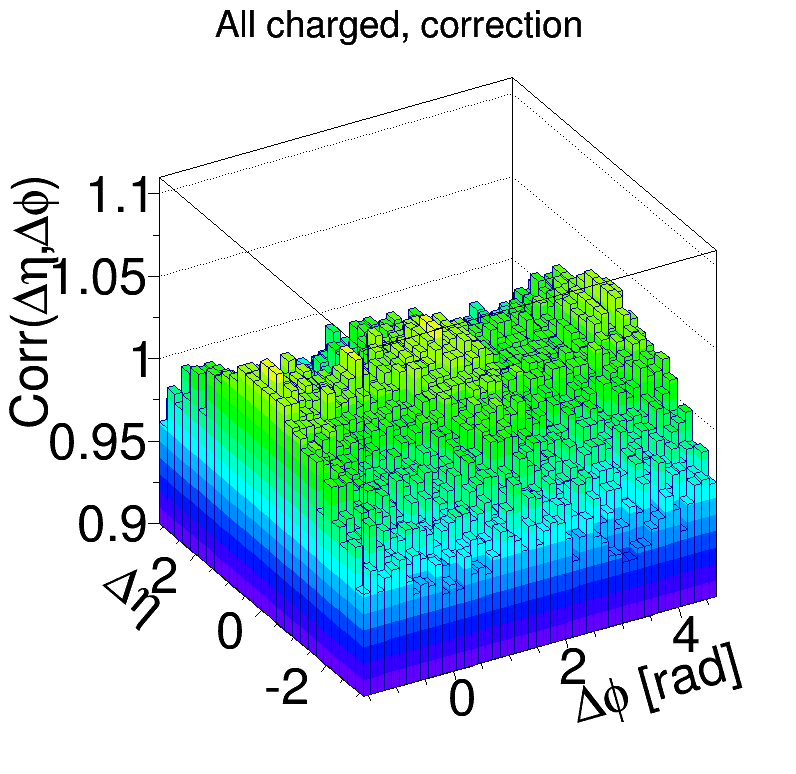}
  \includegraphics[width=0.25\textwidth]{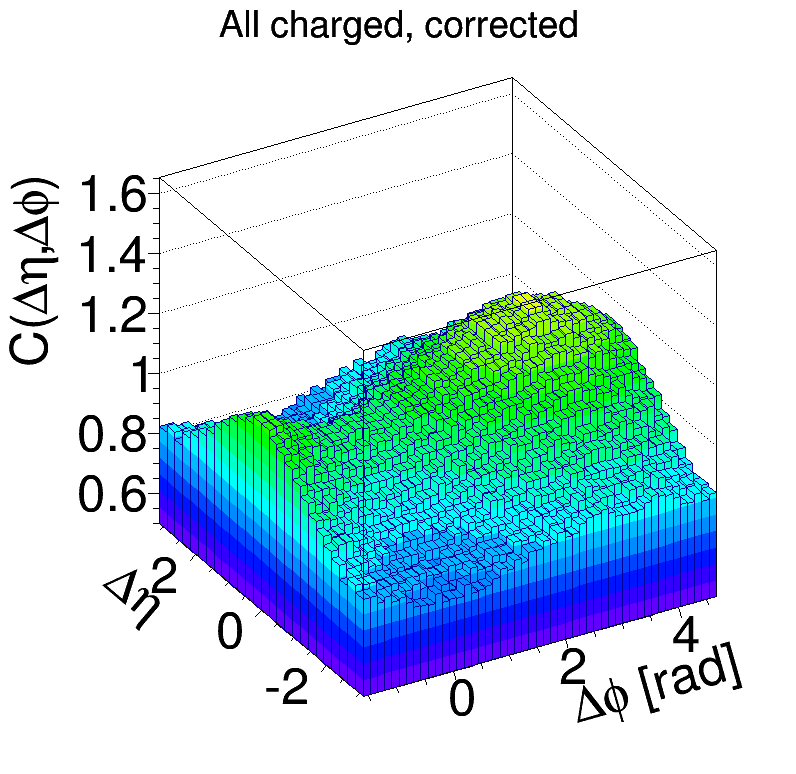}
  
\caption{Example of the correction procedure for the correlation function of p+p interactions at 80~\GeVc beam momentum. Left plot shows correlation function $C(\Delta\eta,\Delta\phi)$ for uncorrected data, middle plot shows the binwise correction factors, right plot presents the corrected correlation function. For all plots the cut $p_T < 1.5$~\GeVc was applied.}
\label{fig:correction_example}
\end{figure}

\section{Corrections}

In order to correct the measured raw correlation function for  possible biases due to trigger and off-line
event selection, track selection, etc. the same analysis was also performed on simulated data. The bin-by-bin correction $Corr(\Delta\eta,\Delta\phi)$ was calculated as the ratio of the correlation functions for generated events from the \Epos~\cite{Werner:2005jf} model (``pure'') and the same events after processing through \Geant detector simulation and reconstruction (``rec''):
\begin{equation}
  \label{eq:corrections}
  Corr(\Delta\eta,\Delta\phi) = \frac{MC_{pure}(\Delta\eta,\Delta\phi)}
                                        {MC_{rec}(\Delta\eta,\Delta\phi)},
\end{equation}
where $MC_{pure}$ is the correlation function obtained for generated events and $MC_{rec}$ is the correlation function for these events after detector simulation and reconstruction. For both ``pure'' and ``rec'' events the \NASixtyOne acceptance was applied.

Corrected correlation functions (example in Fig.~\ref{fig:correction_example}, right) were obtained by multiplying the uncorrected correlation function (Fig.~\ref{fig:correction_example}, left) by the corresponding corrections (Fig.~\ref{fig:correction_example}, middle), namely:

\begin{equation}
\label{eq:corrected_correlations}
  C(\Delta\eta,\Delta\phi) = C^{raw}(\Delta\eta,\Delta\phi)
     \cdot Corr(\Delta\eta,\Delta\phi)
\end{equation}

\section{Results}

\begin{figure}[t]
  \centering
  \includegraphics[width=\textwidth]{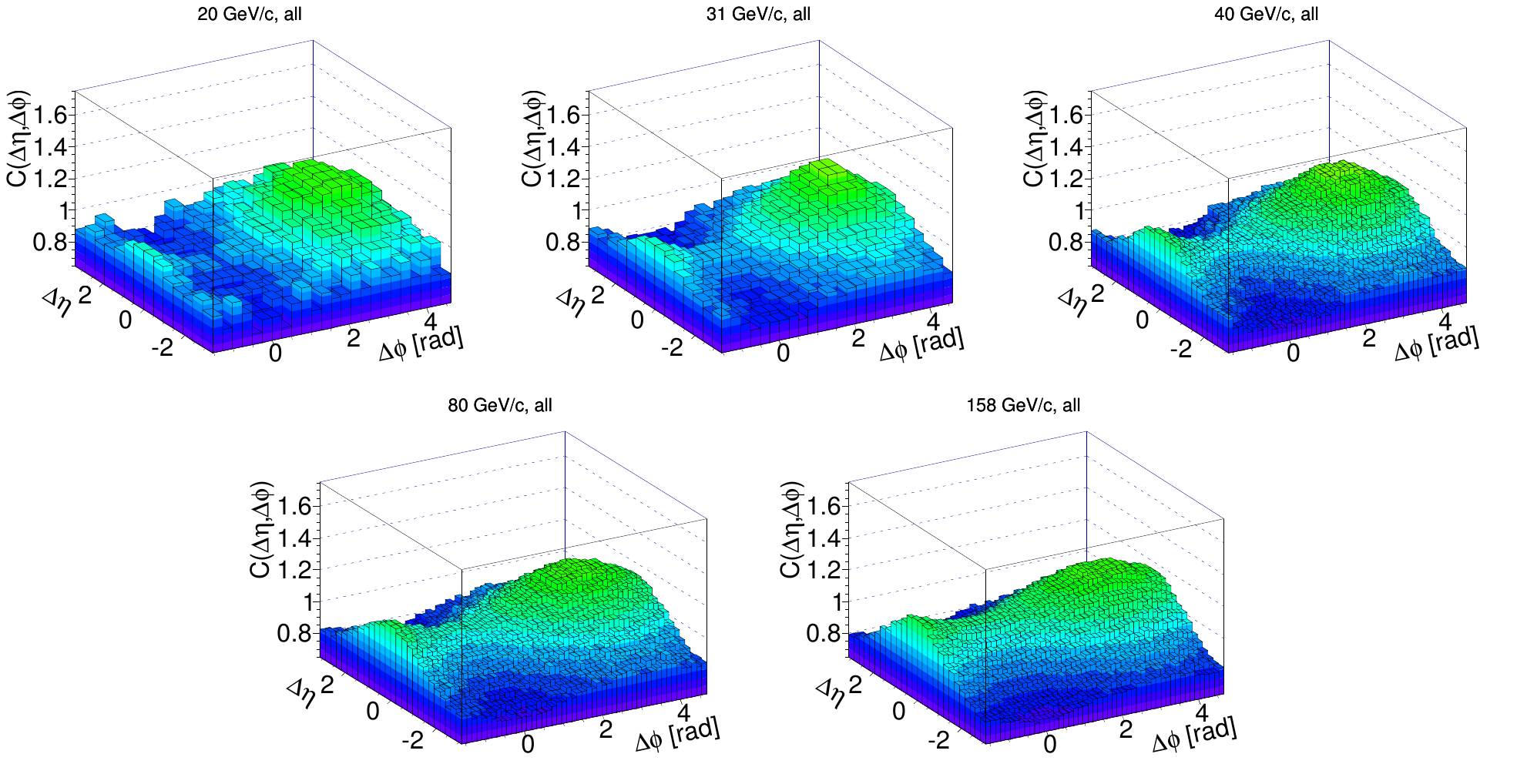}
  \caption{Results on $\Delta\eta\Delta\phi$ correlations for inelastic p+p interactions. Results for all charged particle pairs. The correlation function is mirrored around $(\Delta\eta,\Delta\phi)=(0,0)$.}
  \label{fig:data_corr_all}
\end{figure}

\begin{figure}
  \centering
  \includegraphics[width=\textwidth]{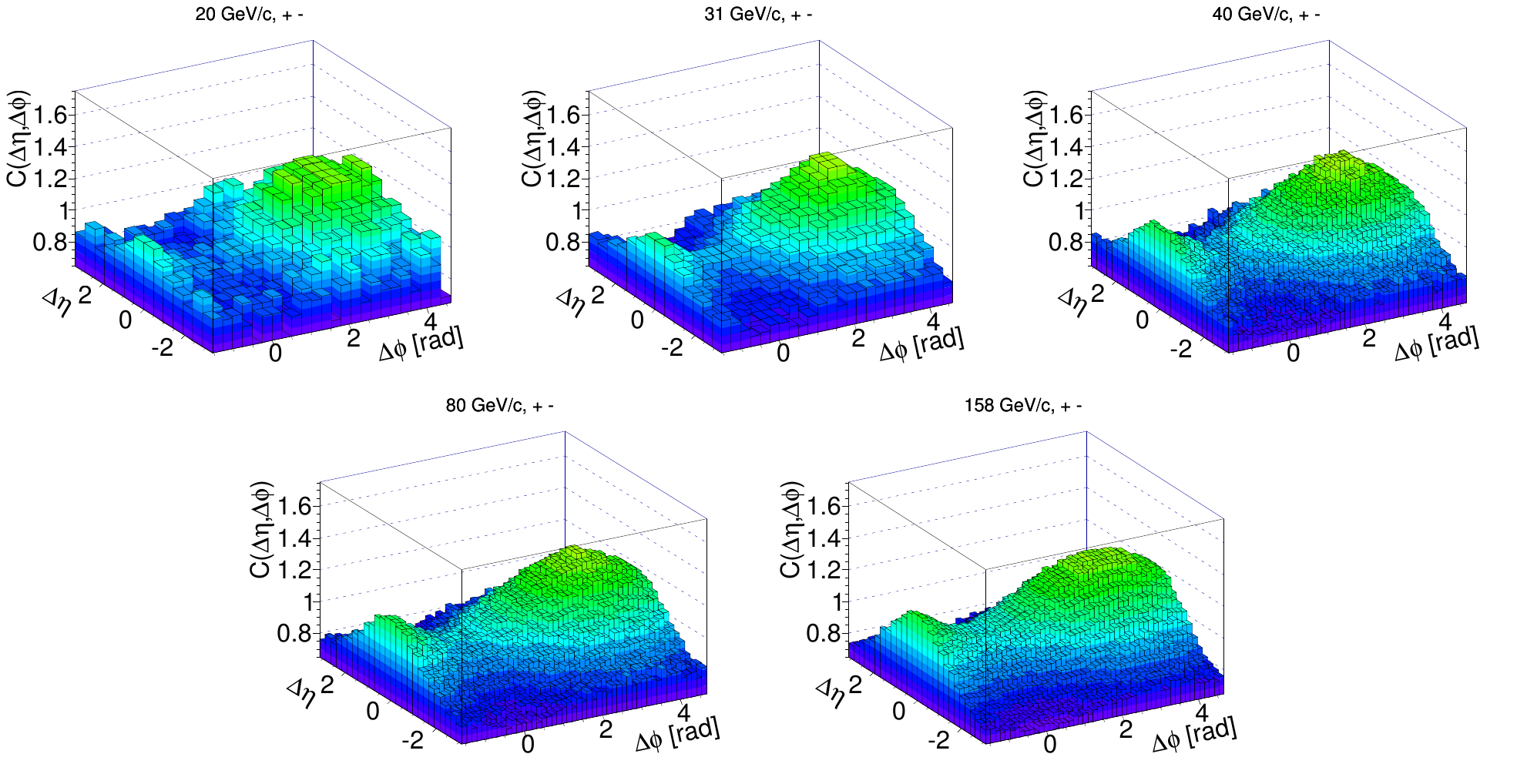}
  \caption{Results on $\Delta\eta\Delta\phi$ correlations for inelastic p+p interactions. Results for unlike-sign pairs. The correlation function is mirrored around $(\Delta\eta,\Delta\phi)=(0,0)$.}
  \label{fig:data_corr_unlike}
\end{figure}

\begin{figure}
  \centering
  \includegraphics[width=\textwidth]{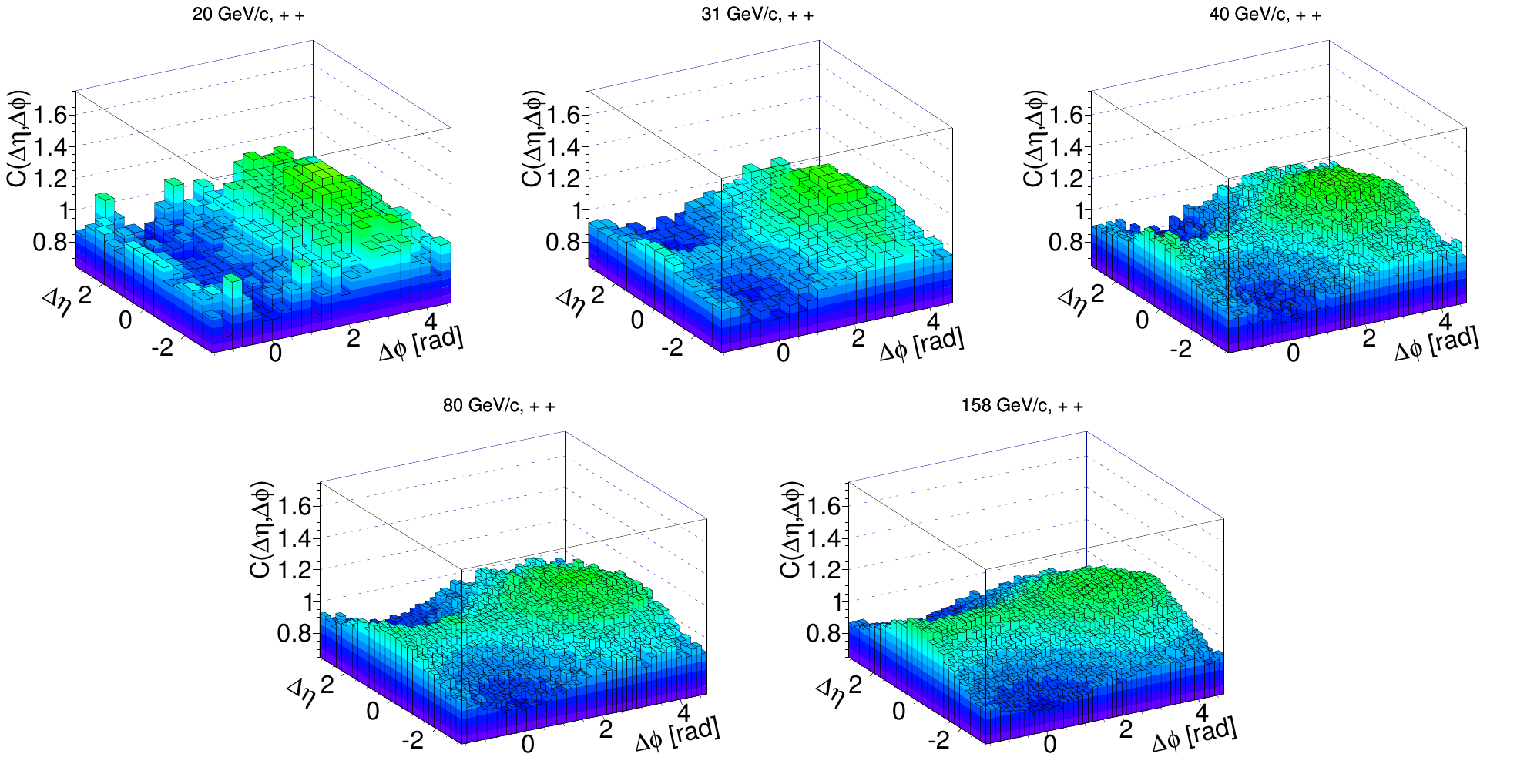}
  \caption{Results on $\Delta\eta\Delta\phi$ correlations for inelastic p+p interactions. Results for positively charged pairs. The correlation function is mirrored around $(\Delta\eta,\Delta\phi)=(0,0)$.}
  \label{fig:data_corr_pos}
\end{figure}

\begin{figure}
  \centering
  \includegraphics[width=\textwidth]{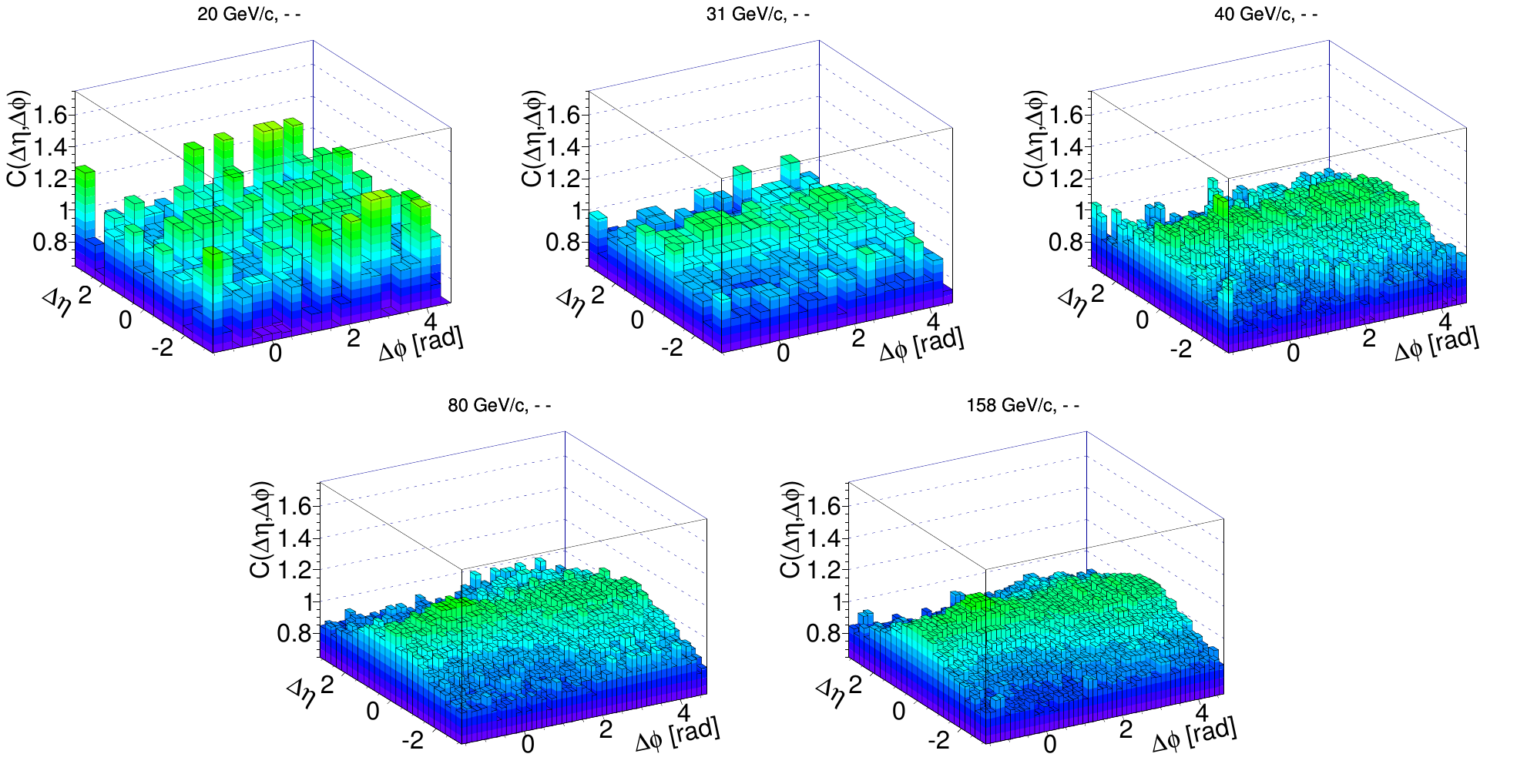}
  \caption{Results on $\Delta\eta\Delta\phi$ correlations for inelastic p+p interactions. Results for negatively charged pairs. The correlation function is mirrored around $(\Delta\eta,\Delta\phi)=(0,0)$.}
  \label{fig:data_corr_neg}
\end{figure}

The corrected correlation functions for all charged pair combinations
(unlike-sign pairs, positively and negatively charged pairs)
are presented in Figs.~\ref{fig:data_corr_all}, \ref{fig:data_corr_unlike},
\ref{fig:data_corr_pos} and \ref{fig:data_corr_neg}, respectively.

Several structures can be seen in the plots:
\begin{itemize}
\item A maximum at $(\Delta\eta,\Delta\phi) = (0,\pi)$, probably a result of resonance decays and momentum conservation. It is strongest for unlike-sign pairs and significantly weaker for same charge pairs.
\item Weak enhancement at $(\Delta\eta,\Delta\phi) = (0,0)$, likely due to Coulomb interactions (unlike-sign pairs) and quantum statistics (same charge pairs).
\end{itemize}

\subsection{Comparison with the EPOS model}

\begin{figure}
  \centering
  \includegraphics[width=0.3\textwidth]{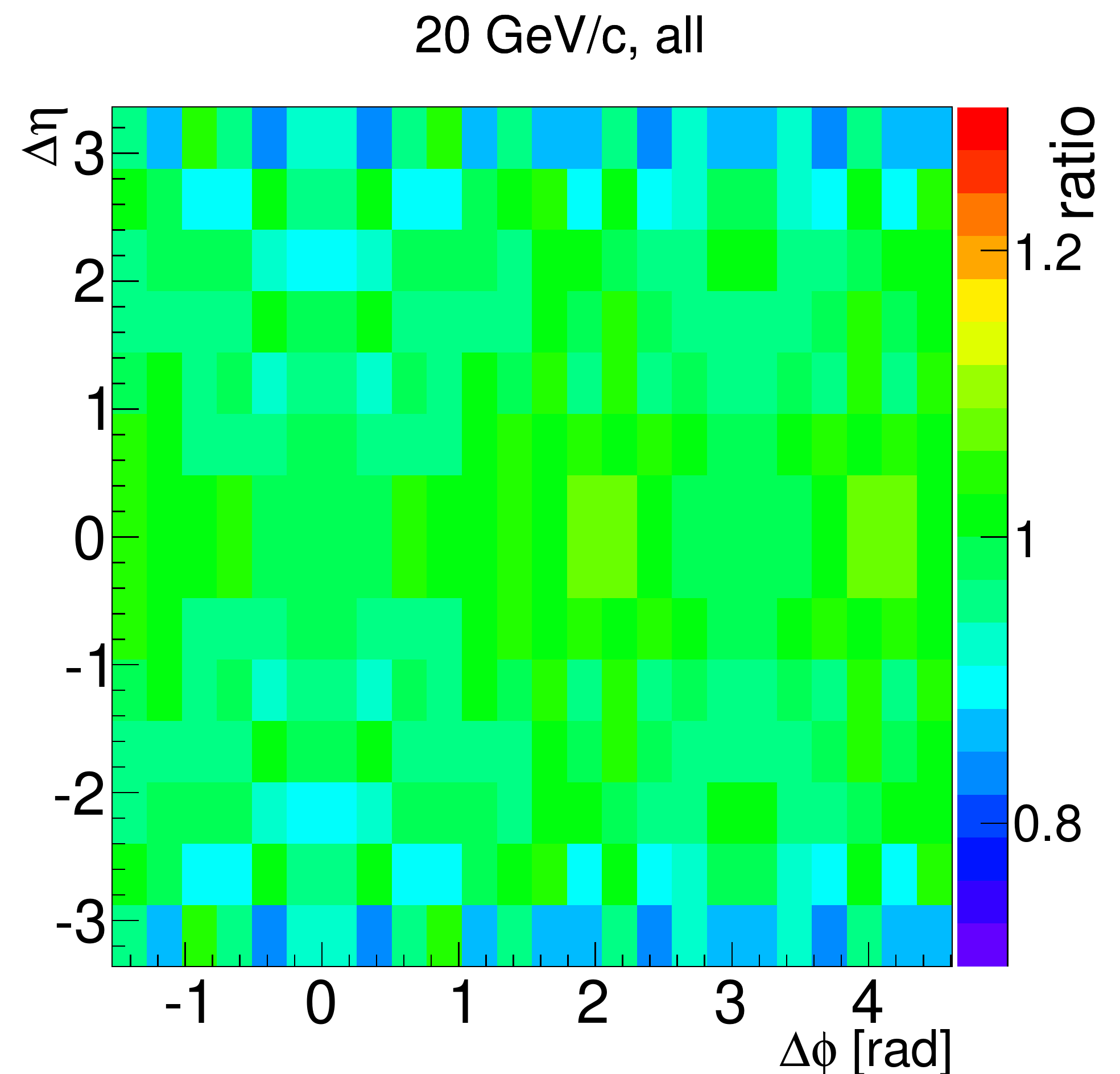}
  \includegraphics[width=0.3\textwidth]{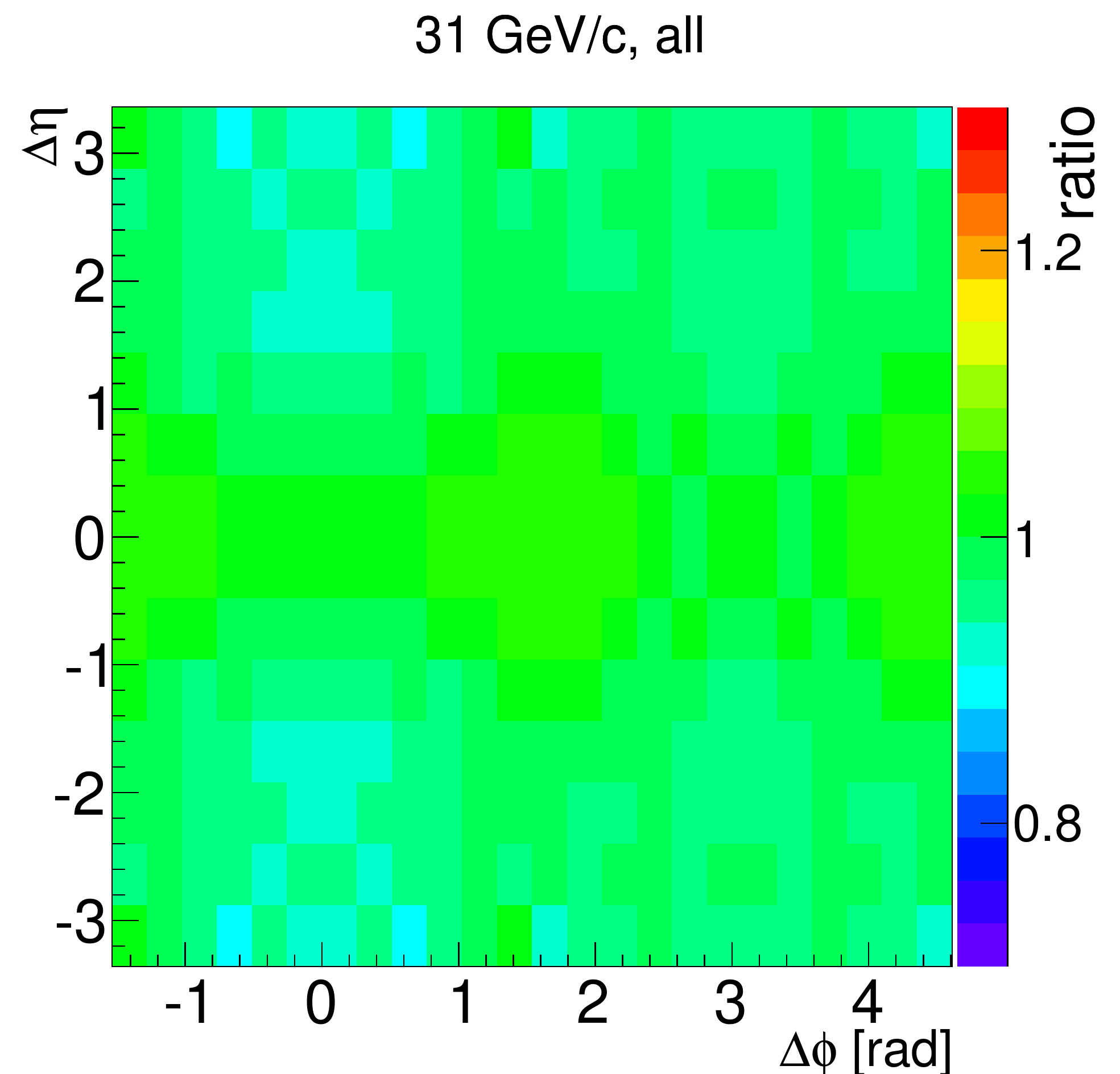}
  \includegraphics[width=0.3\textwidth]{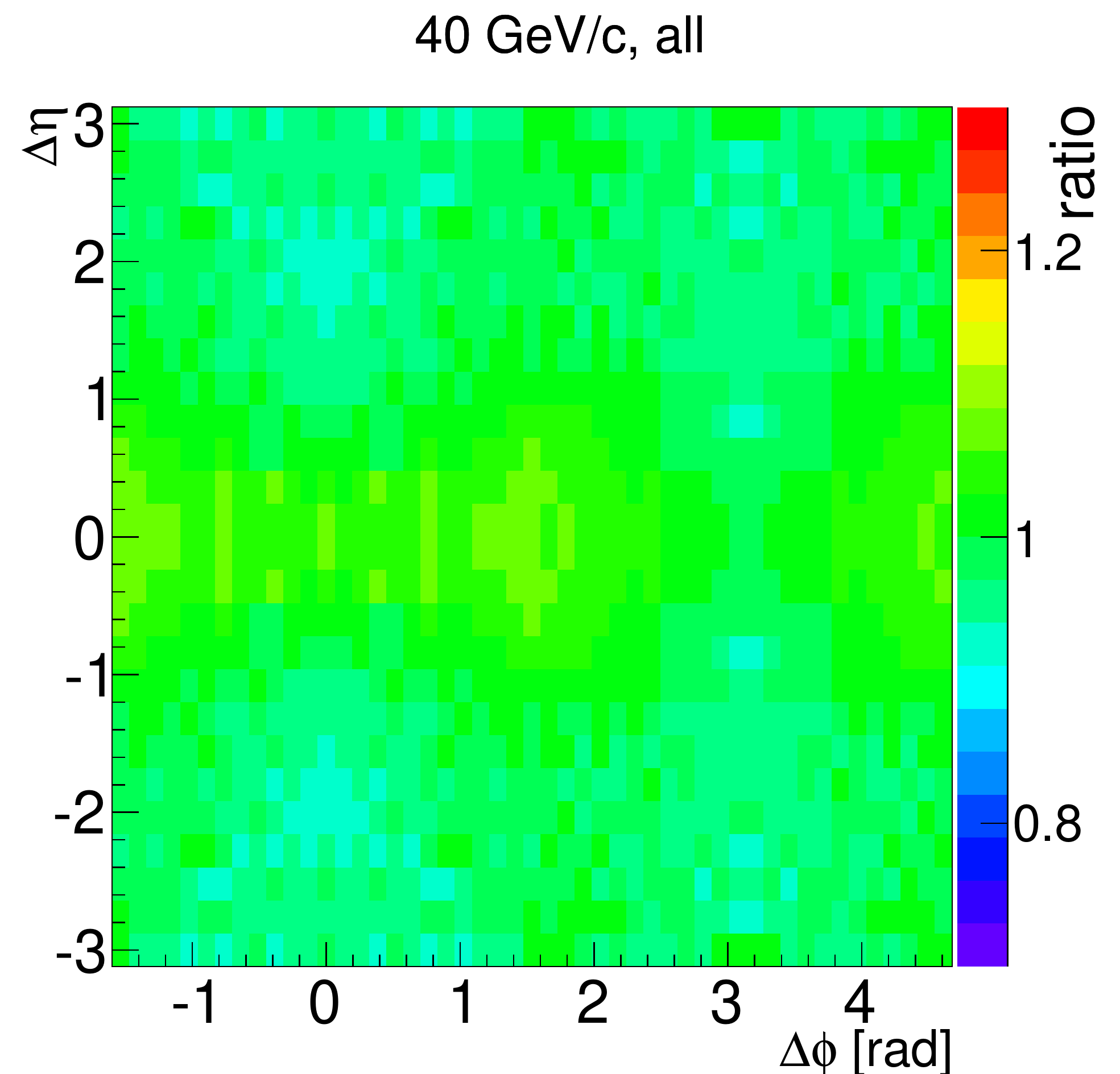}
\\
  \includegraphics[width=0.3\textwidth]{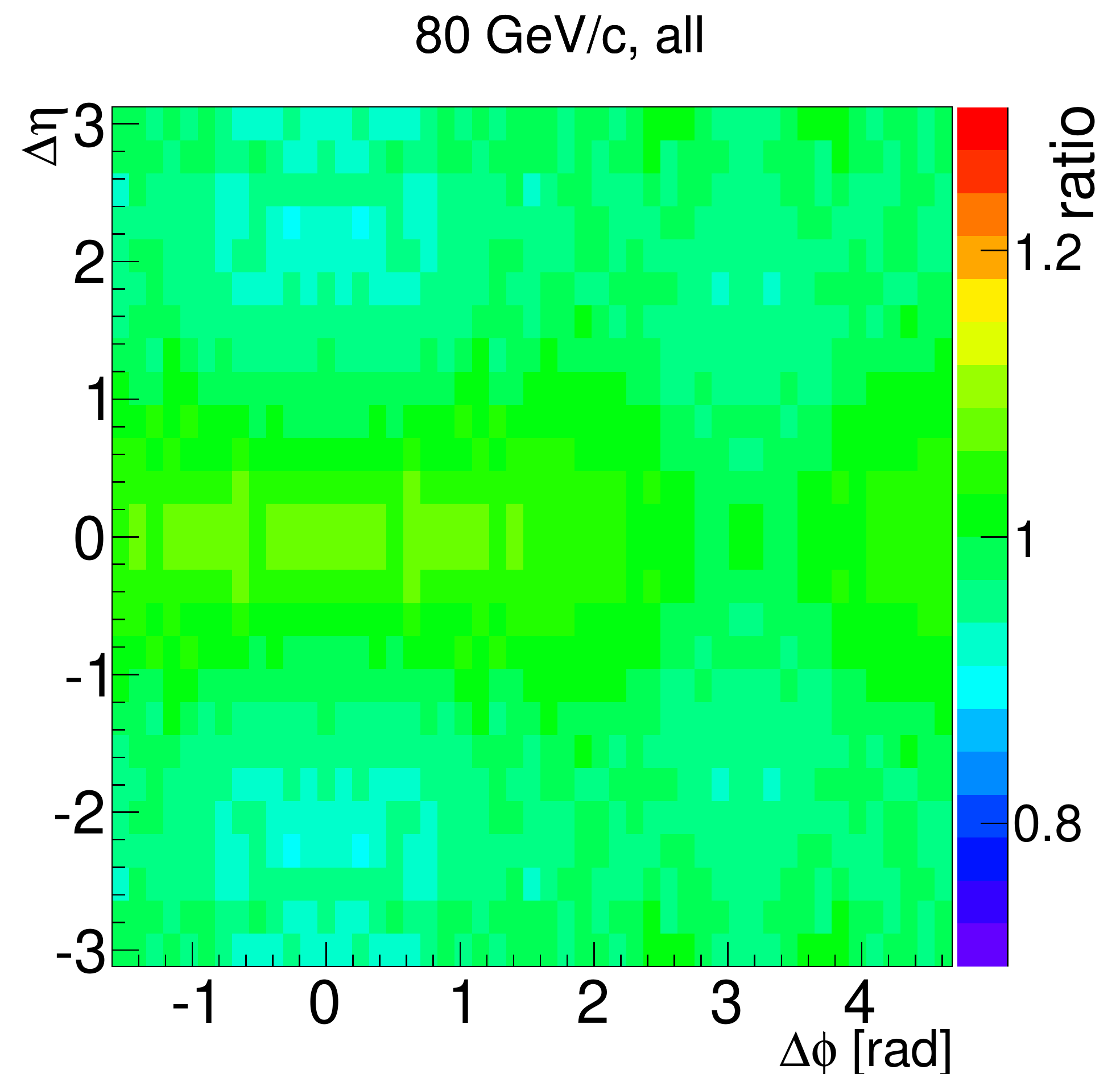}
  \includegraphics[width=0.3\textwidth]{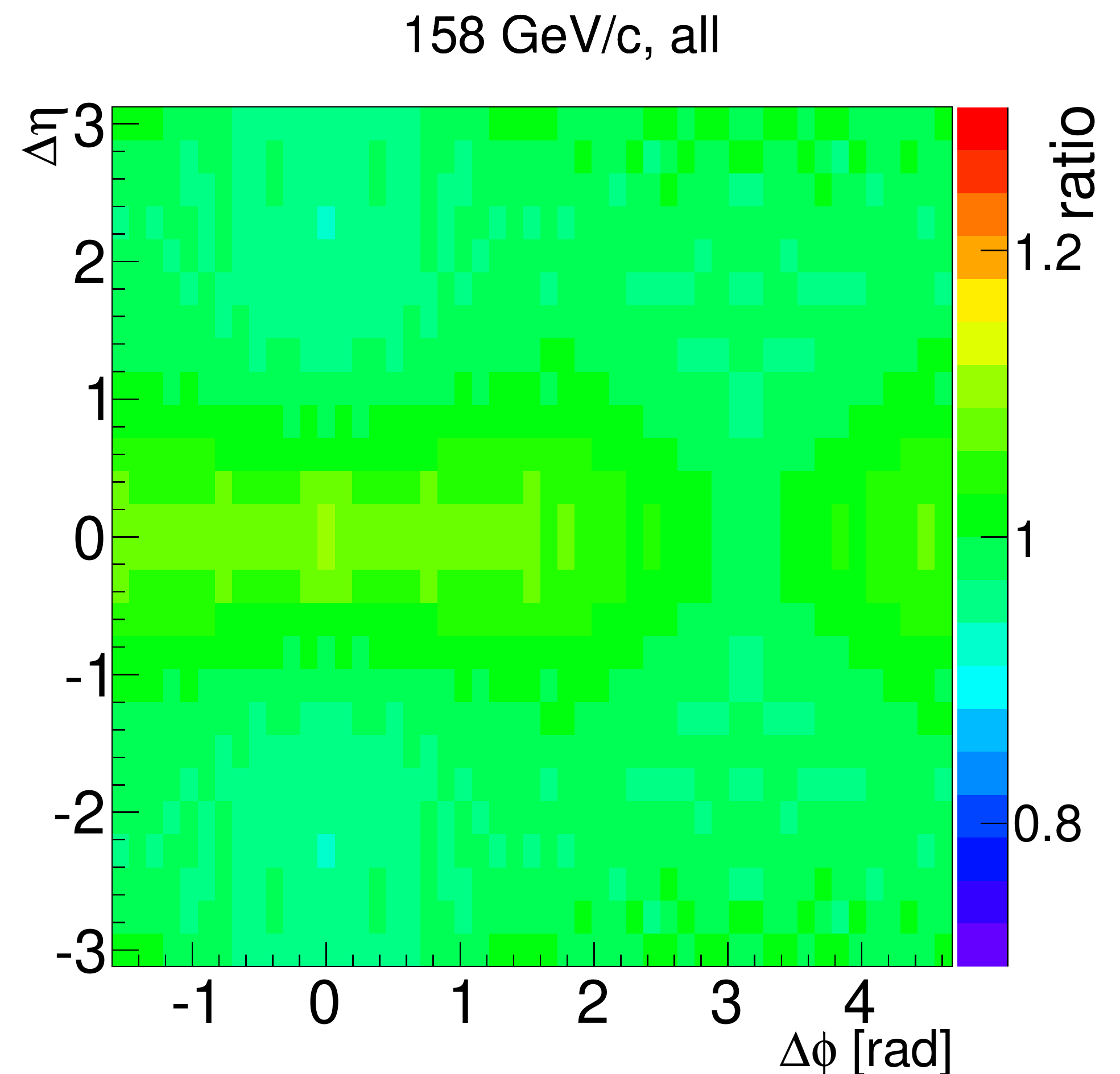}
  \caption{The ratio of $C(\Delta\eta,\Delta\phi)$ for data and \Epos.
All charged particle pairs.
The correlation function is mirrored around $(\Delta\eta,\Delta\phi)=(0,0)$.}
  \label{fig:data_vs_mc_ratio_all}
\end{figure}

\begin{figure}
  \centering
  \includegraphics[width=0.3\textwidth]{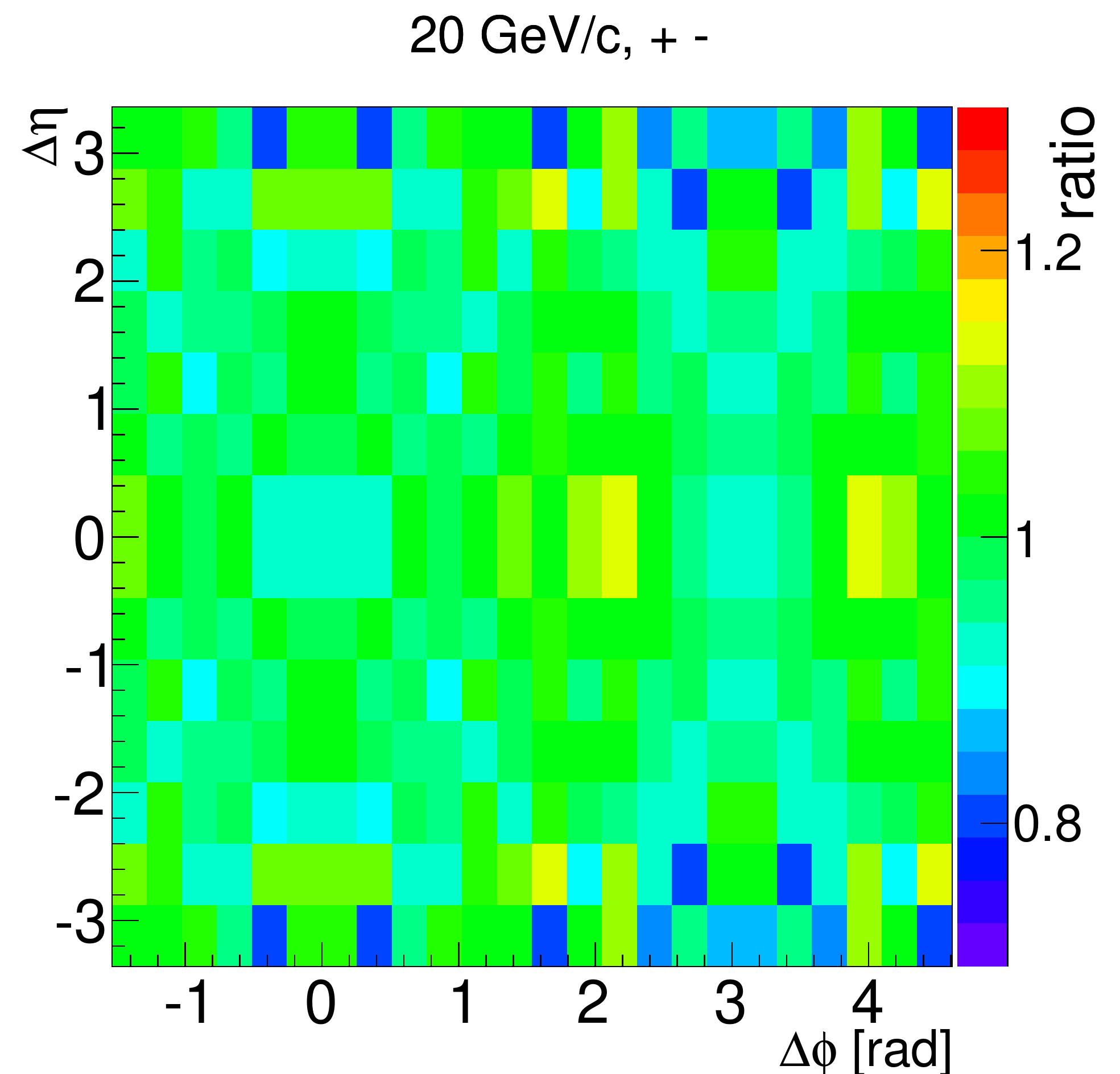}
  \includegraphics[width=0.3\textwidth]{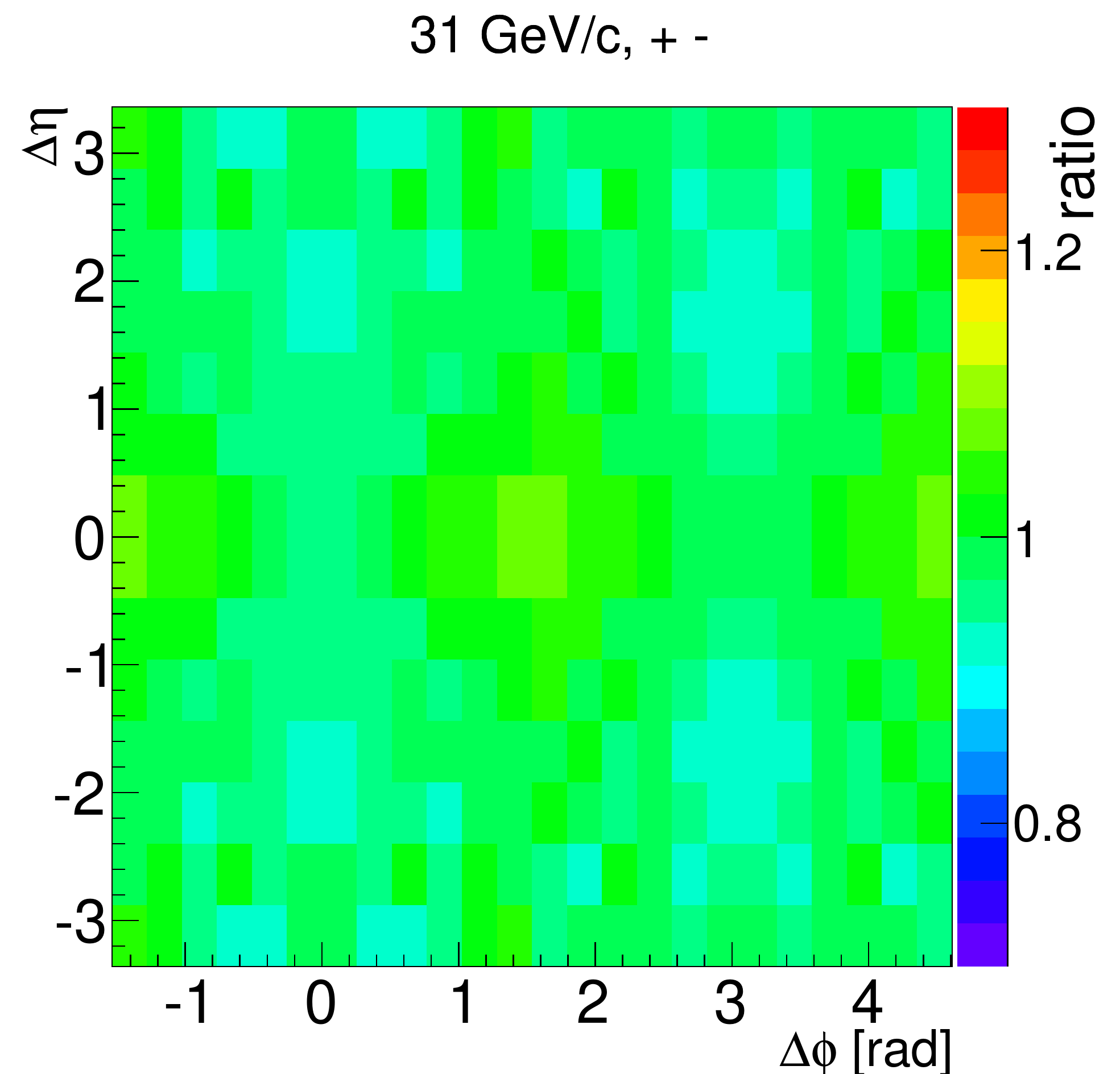}
  \includegraphics[width=0.3\textwidth]{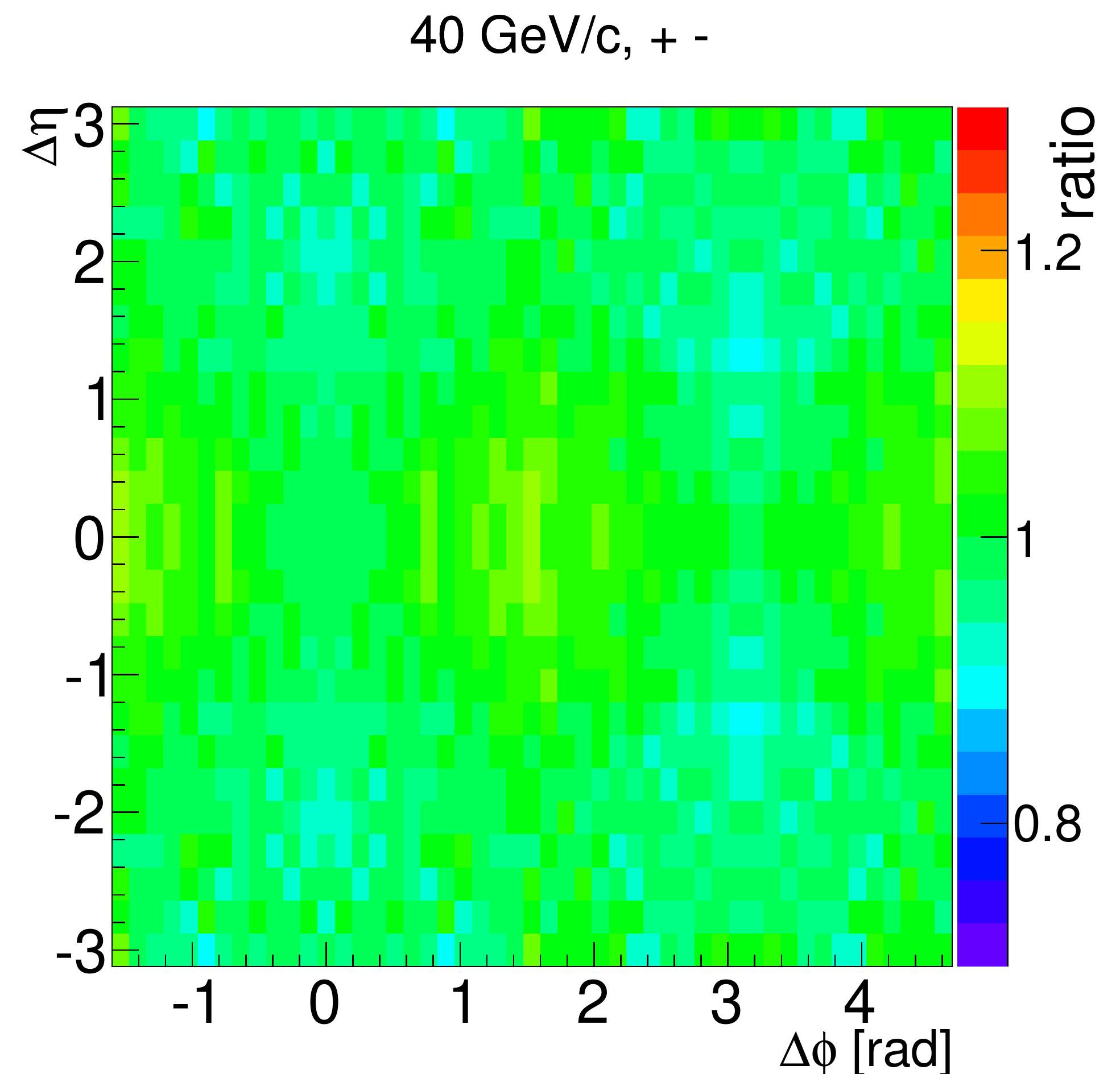}
\\
  \includegraphics[width=0.3\textwidth]{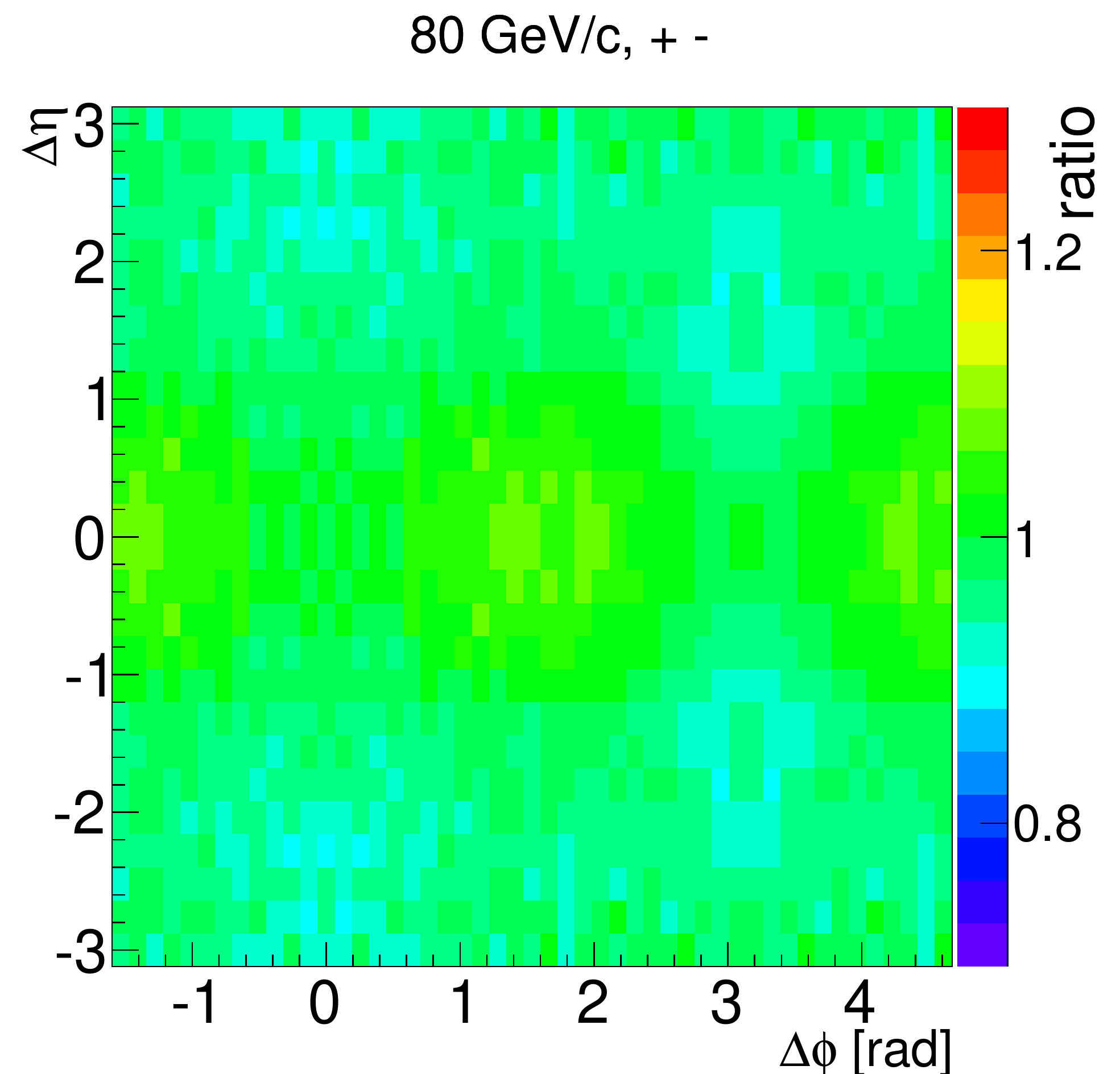}
  \includegraphics[width=0.3\textwidth]{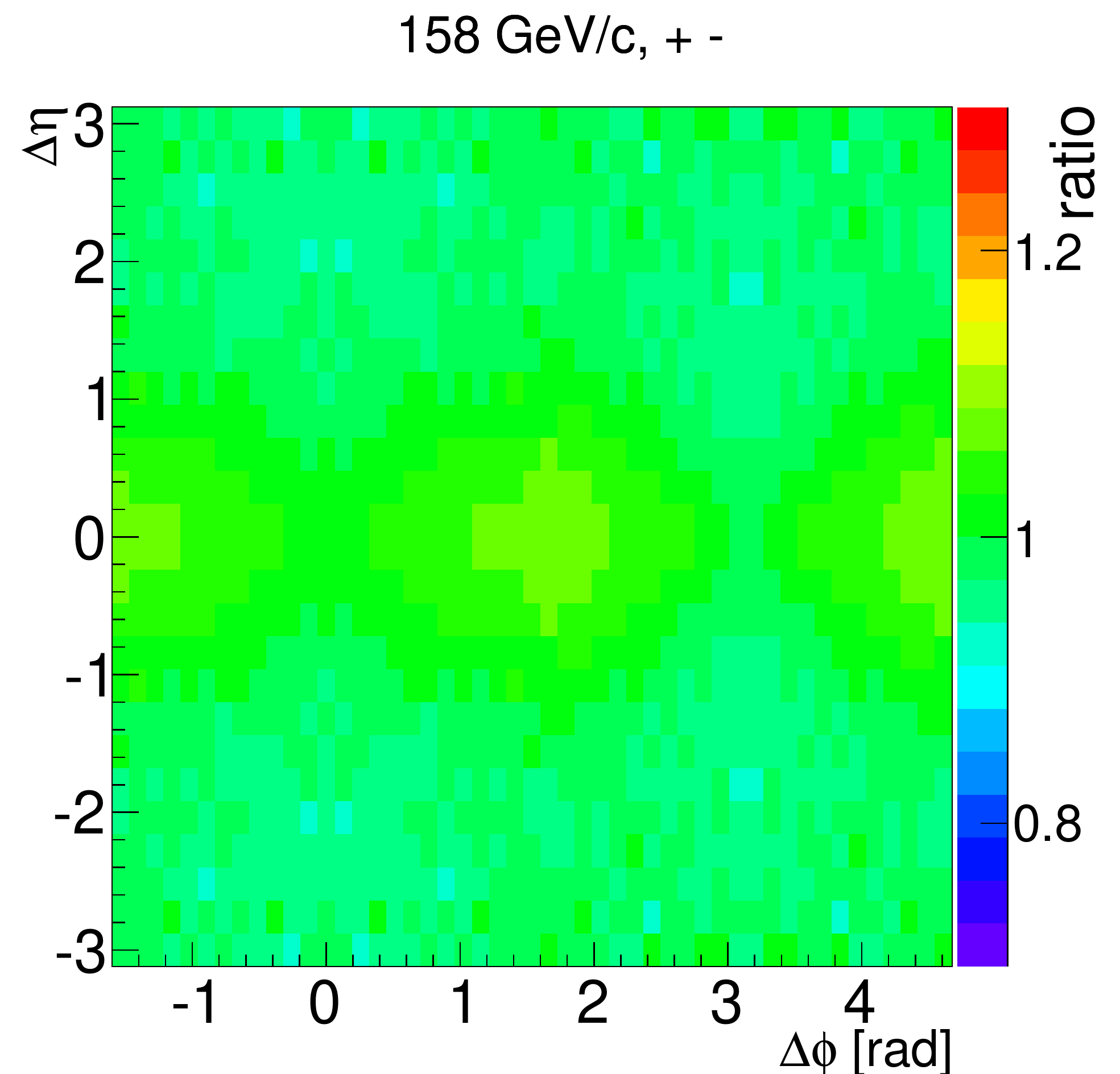}
  \caption{The ratio of $C(\Delta\eta,\Delta\phi)$ for data and \Epos.
  Unlike-sign pairs. The correlation function is mirrored around $(\Delta\eta,\Delta\phi)=(0,0)$.}
  \label{fig:data_vs_mc_ratio_unlike}
\end{figure}

\begin{figure}[h]
  \centering
  \includegraphics[width=0.3\textwidth]{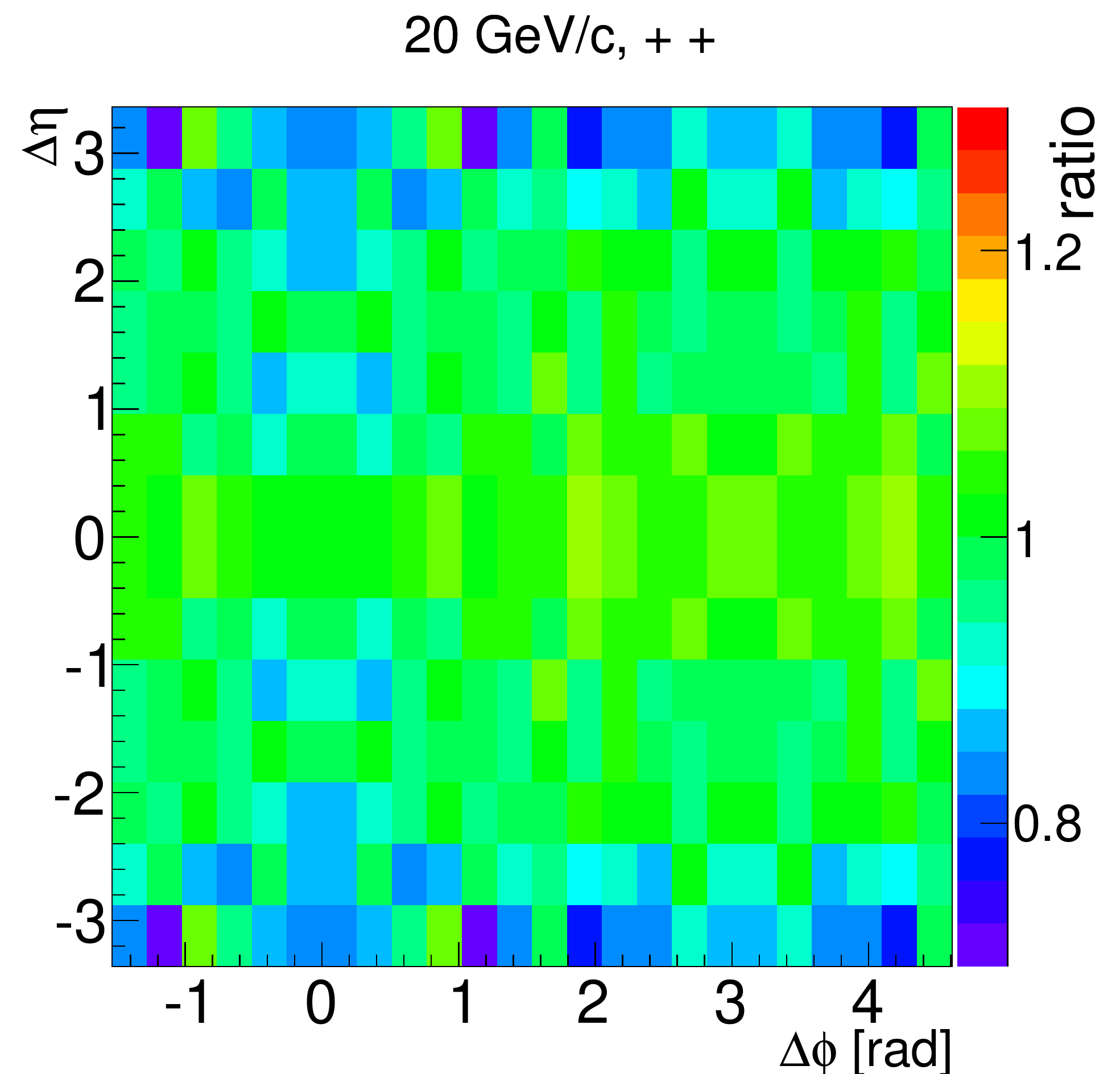}
  \includegraphics[width=0.3\textwidth]{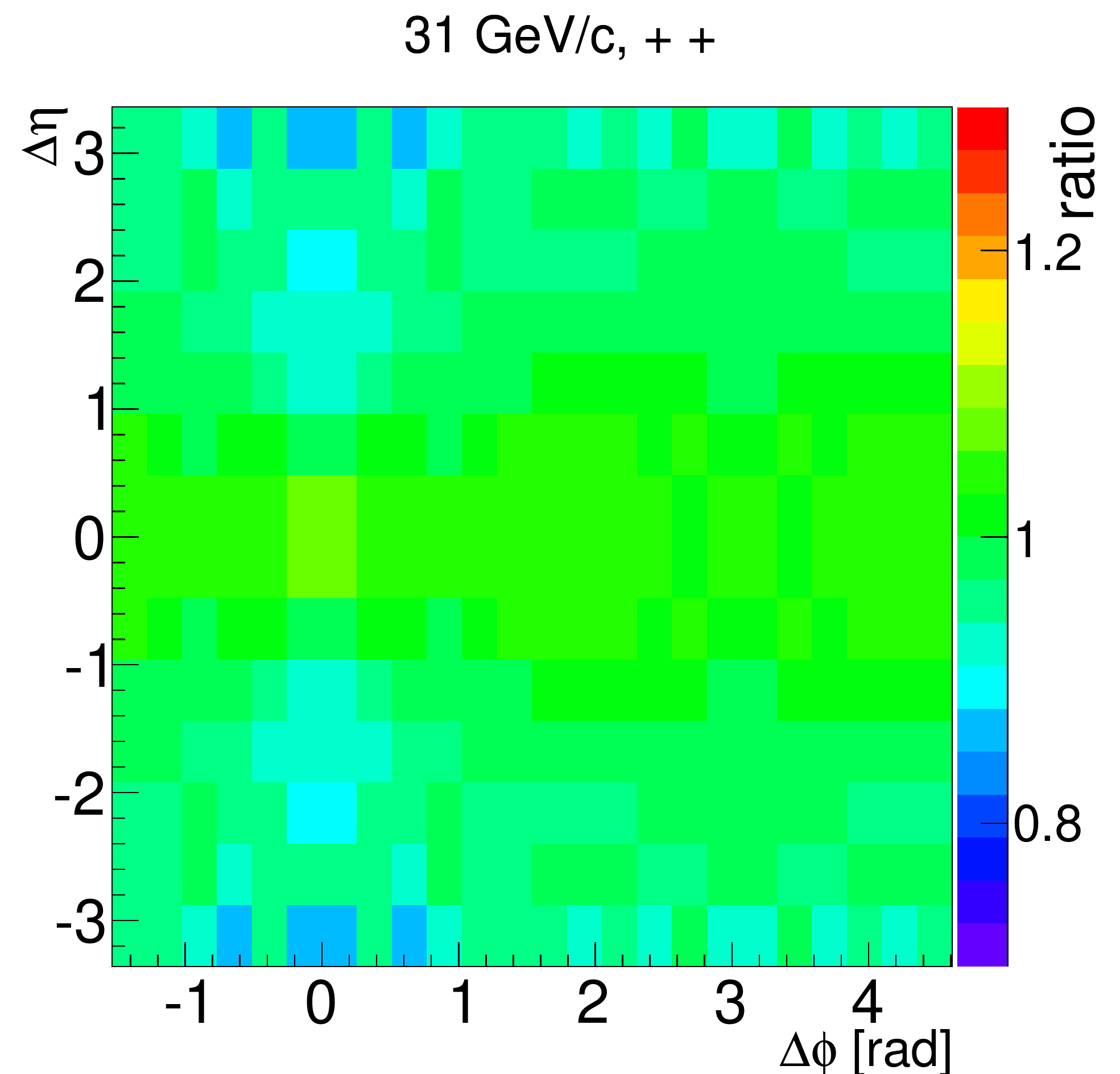}
  \includegraphics[width=0.3\textwidth]{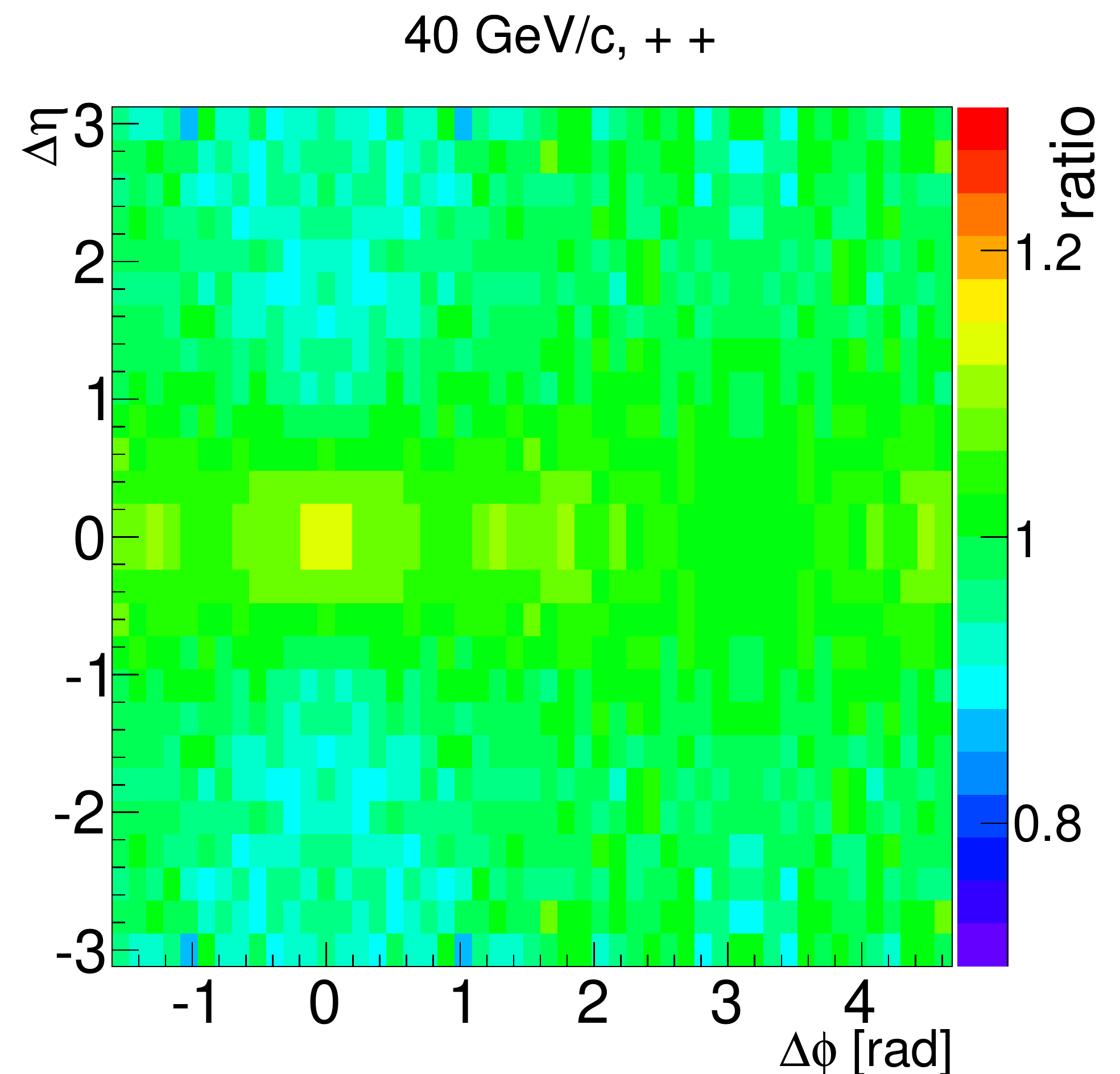}
\\
  \includegraphics[width=0.3\textwidth]{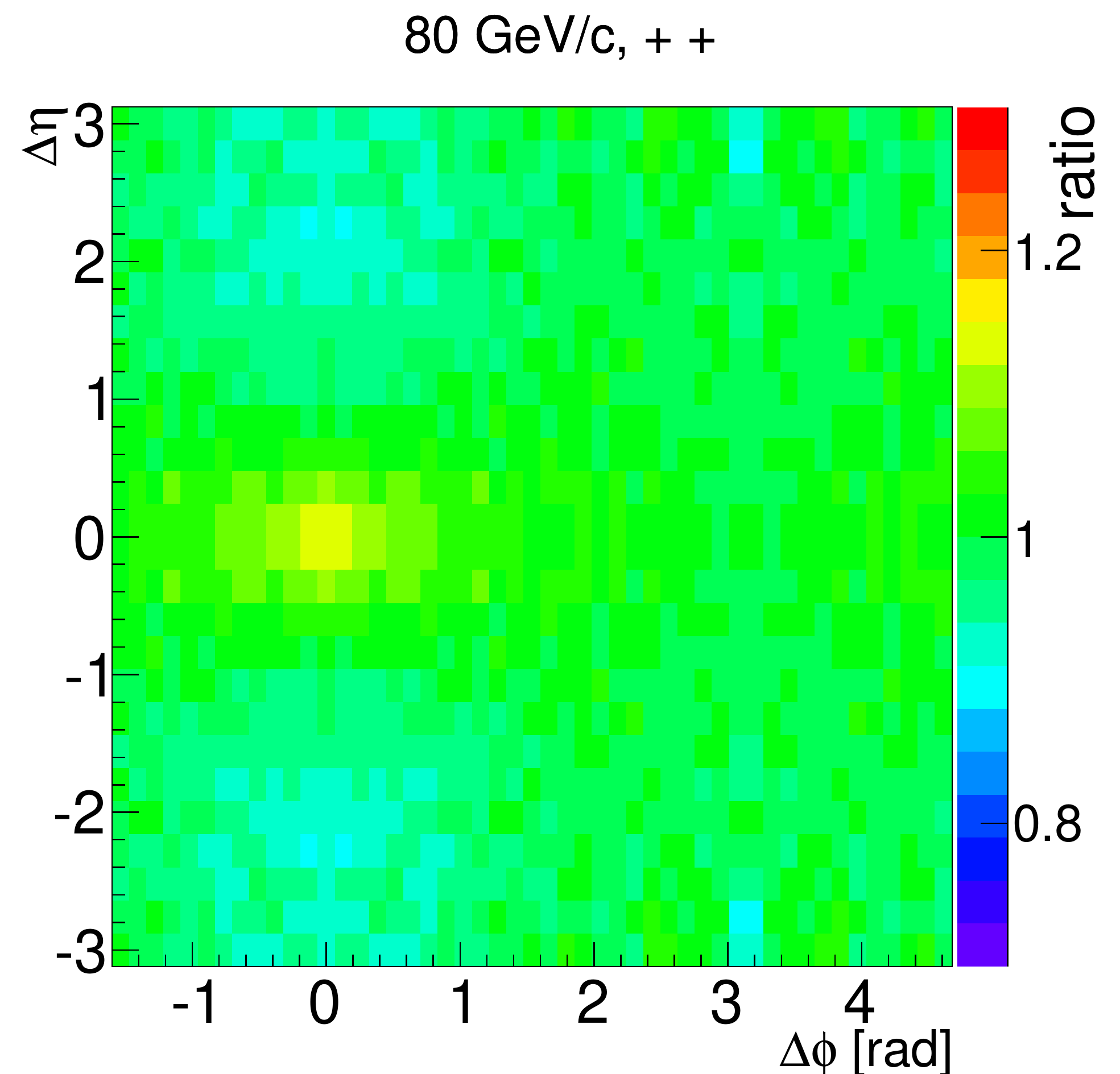}
  \includegraphics[width=0.3\textwidth]{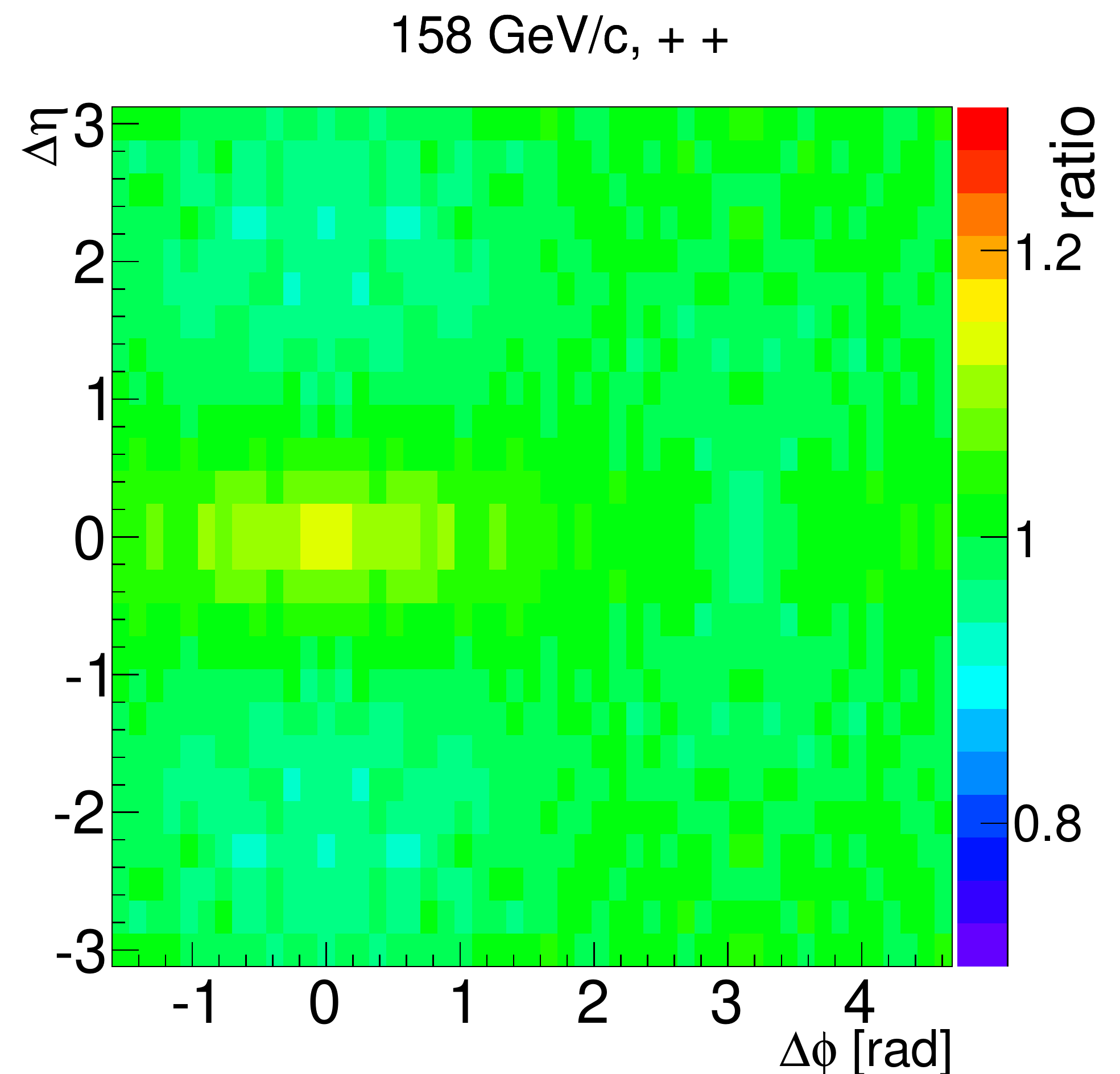}
  \caption{The ratio of $C(\Delta\eta,\Delta\phi)$ for data and \Epos.
  Positively charged pairs. The correlation function is mirrored around $(\Delta\eta,\Delta\phi)=(0,0)$.}
  \label{fig:data_vs_mc_ratio_pos}
\end{figure}

\begin{figure}
  \centering
  \includegraphics[width=0.3\textwidth]{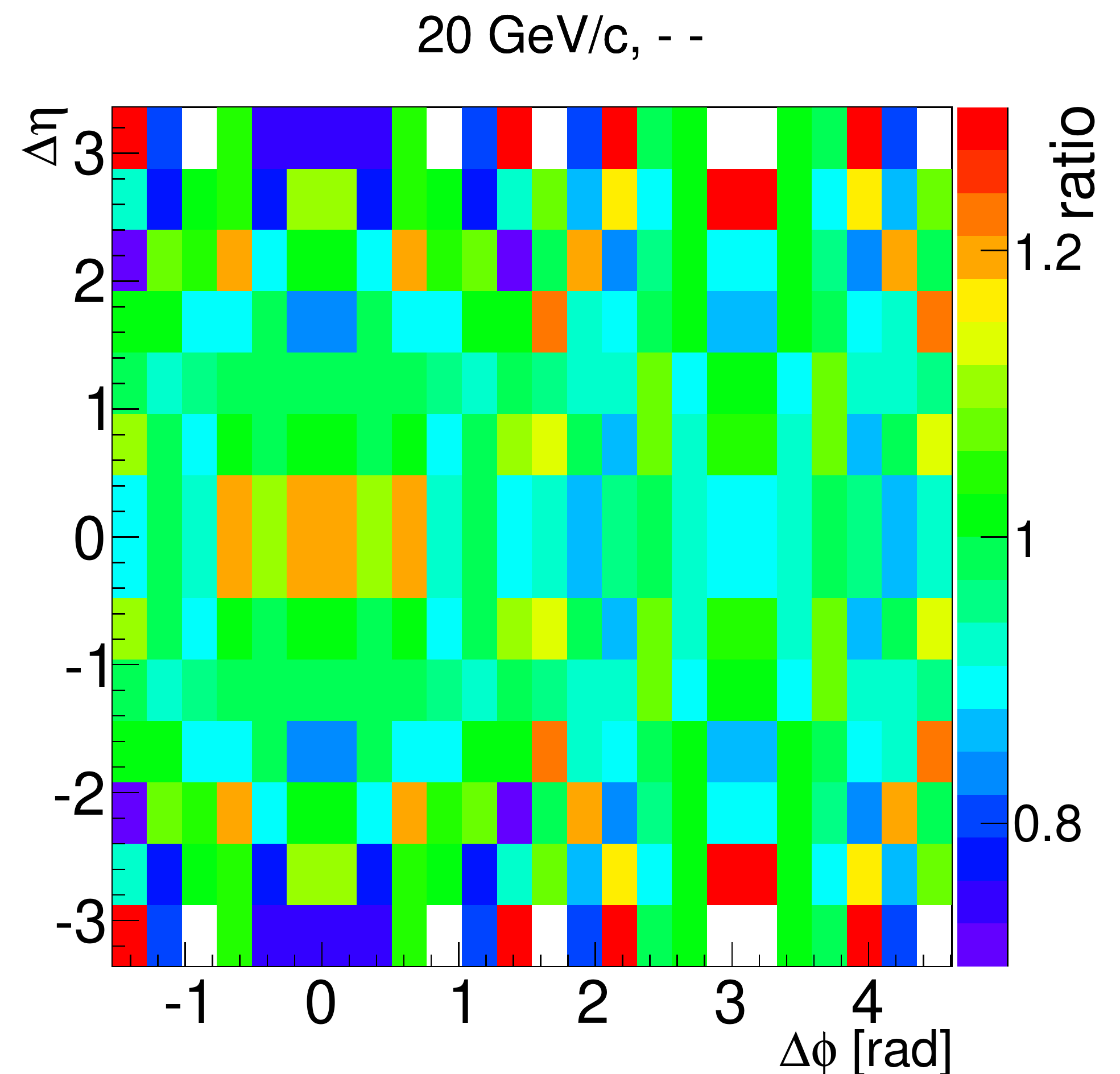}
  \includegraphics[width=0.3\textwidth]{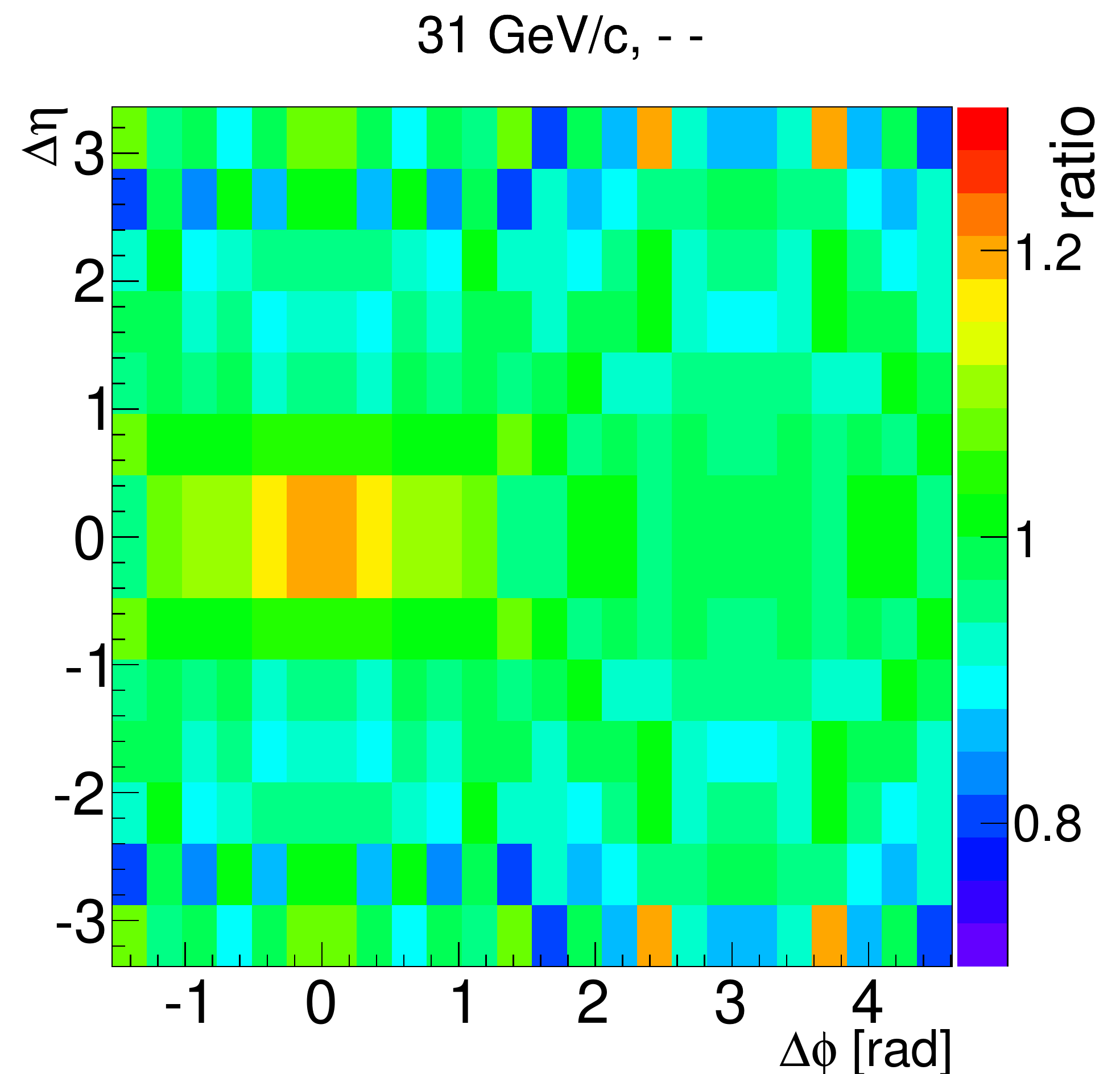}
  \includegraphics[width=0.3\textwidth]{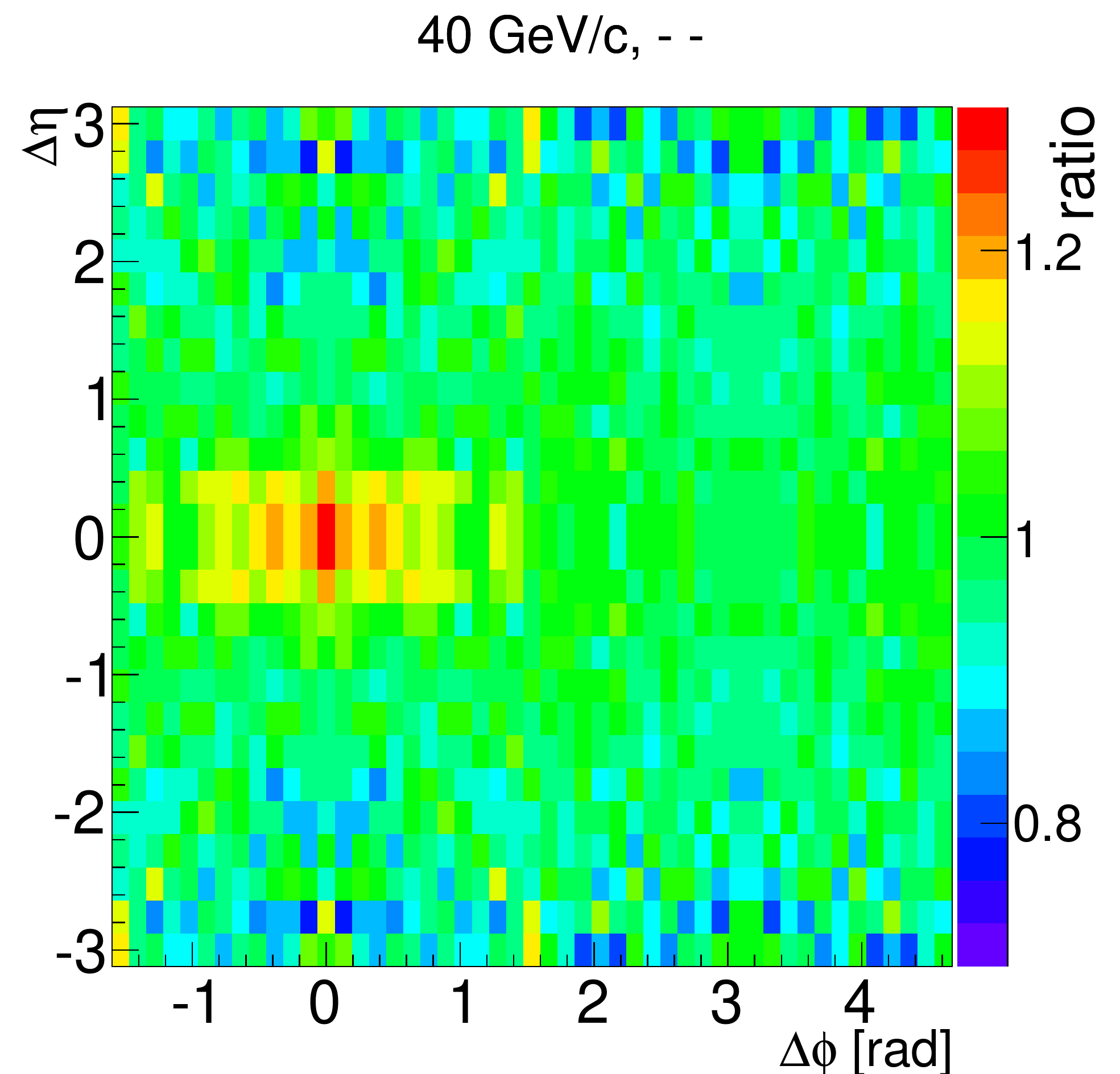}
\\
  \includegraphics[width=0.3\textwidth]{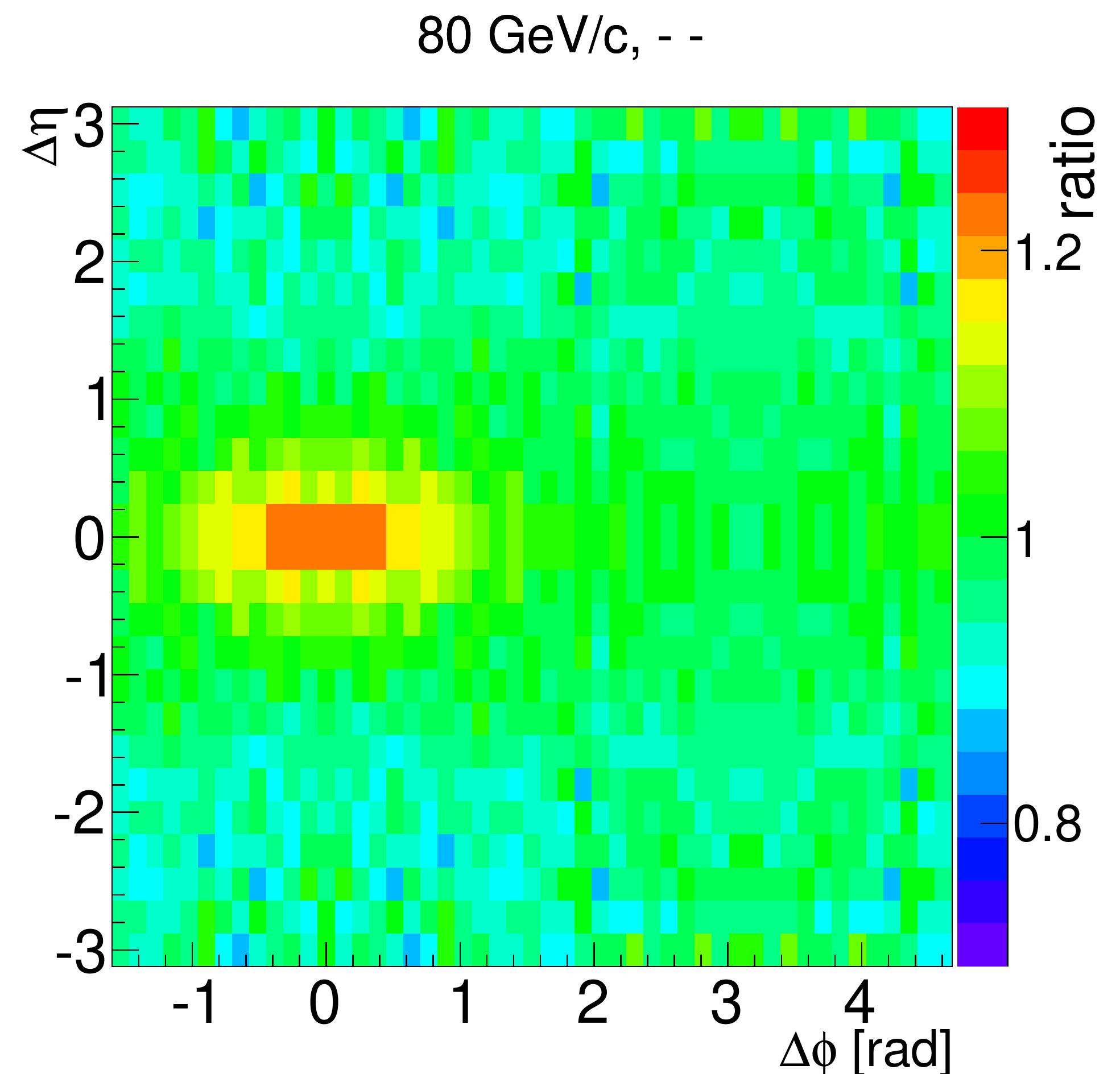}
  \includegraphics[width=0.3\textwidth]{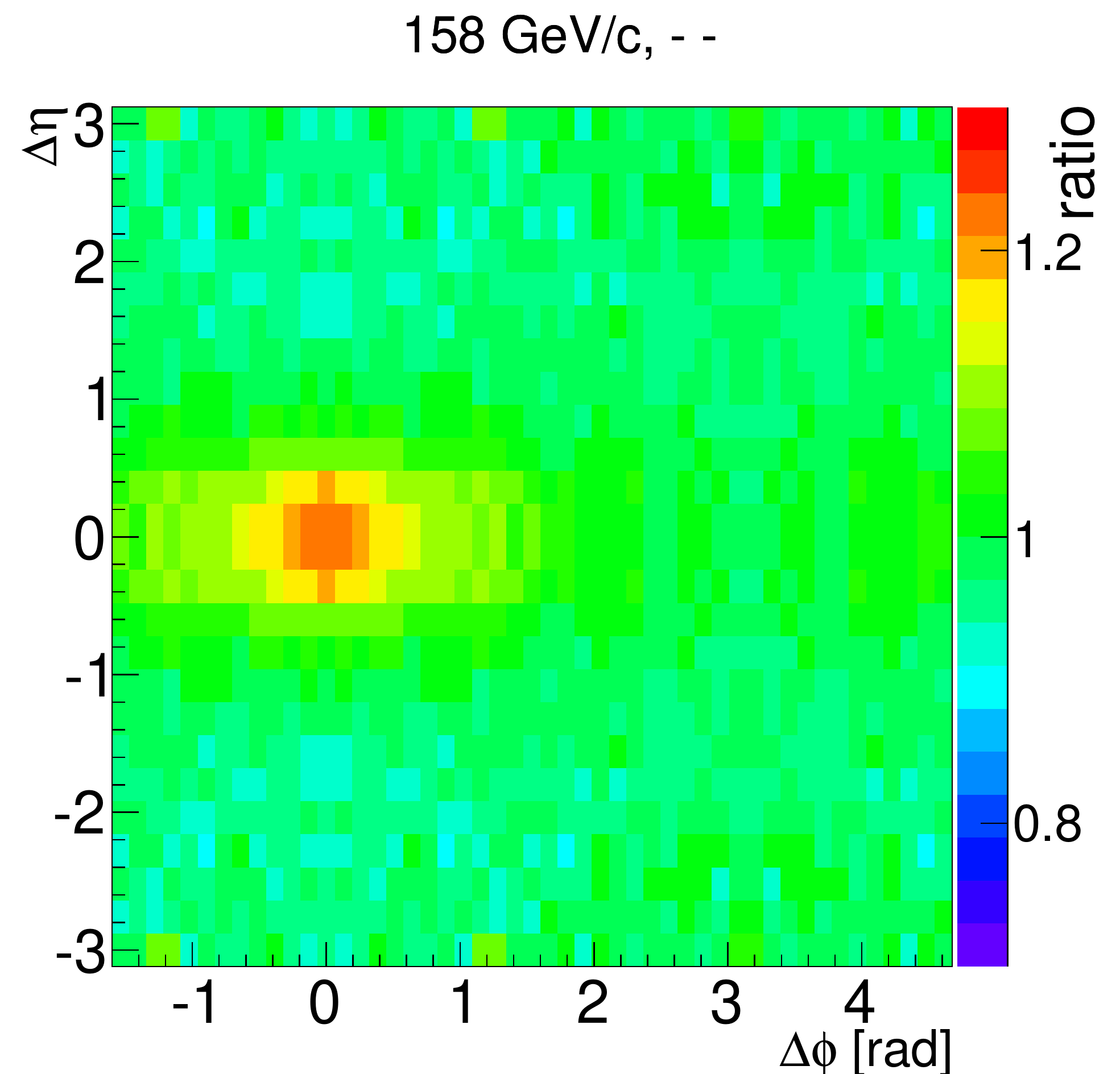}
  \caption{The ratio of $C(\Delta\eta,\Delta\phi)$ for data and \Epos.
  Negatively charged pairs. The correlation function is mirrored around $(\Delta\eta,\Delta\phi)=(0,0)$.}
  \label{fig:data_vs_mc_ratio_neg}
\end{figure}

The ratio of the experimental results and predictions of the \Epos model is shown in Figs.~\ref{fig:data_vs_mc_ratio_all}, \ref{fig:data_vs_mc_ratio_unlike}, \ref{fig:data_vs_mc_ratio_pos} and \ref{fig:data_vs_mc_ratio_neg}. In general \Epos reproduces results from p+p interactions well.
However, the model does not reproduce the $(\Delta\eta,\Delta\phi)=(0,0)$ peak. This can be seen for positively and negatively charged pairs (Figs.~\ref{fig:data_vs_mc_ratio_pos} and \ref{fig:data_vs_mc_ratio_neg} respectively). Note, that \Epos does not include Bose-Einstein correlations and Coulomb interactions.

\clearpage

\subsection{Influence of $p_T$ cut at 158\GeVc}

\begin{figure}
  \centering
  \includegraphics[width=0.45\textwidth]{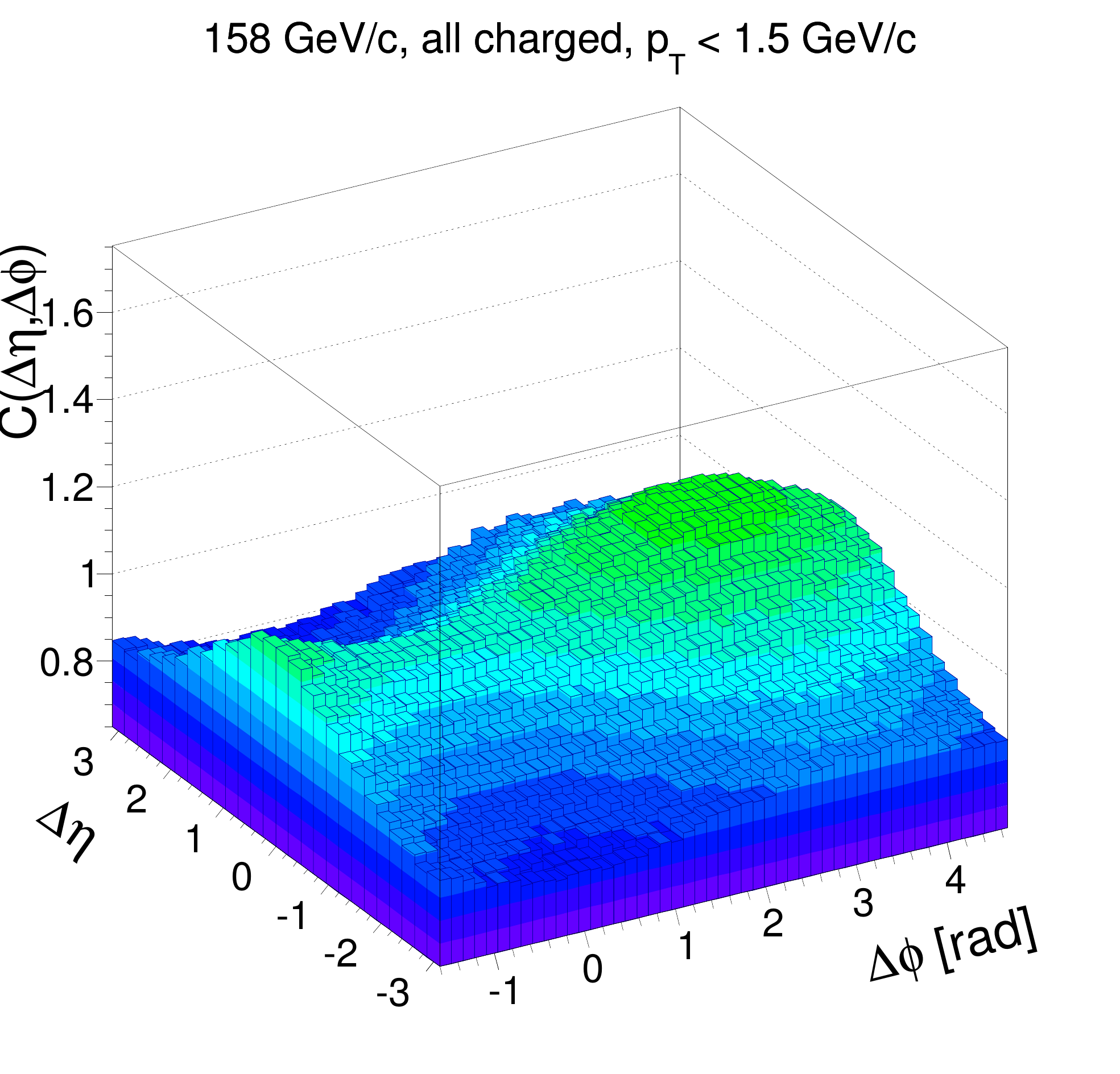}
  \includegraphics[width=0.45\textwidth]{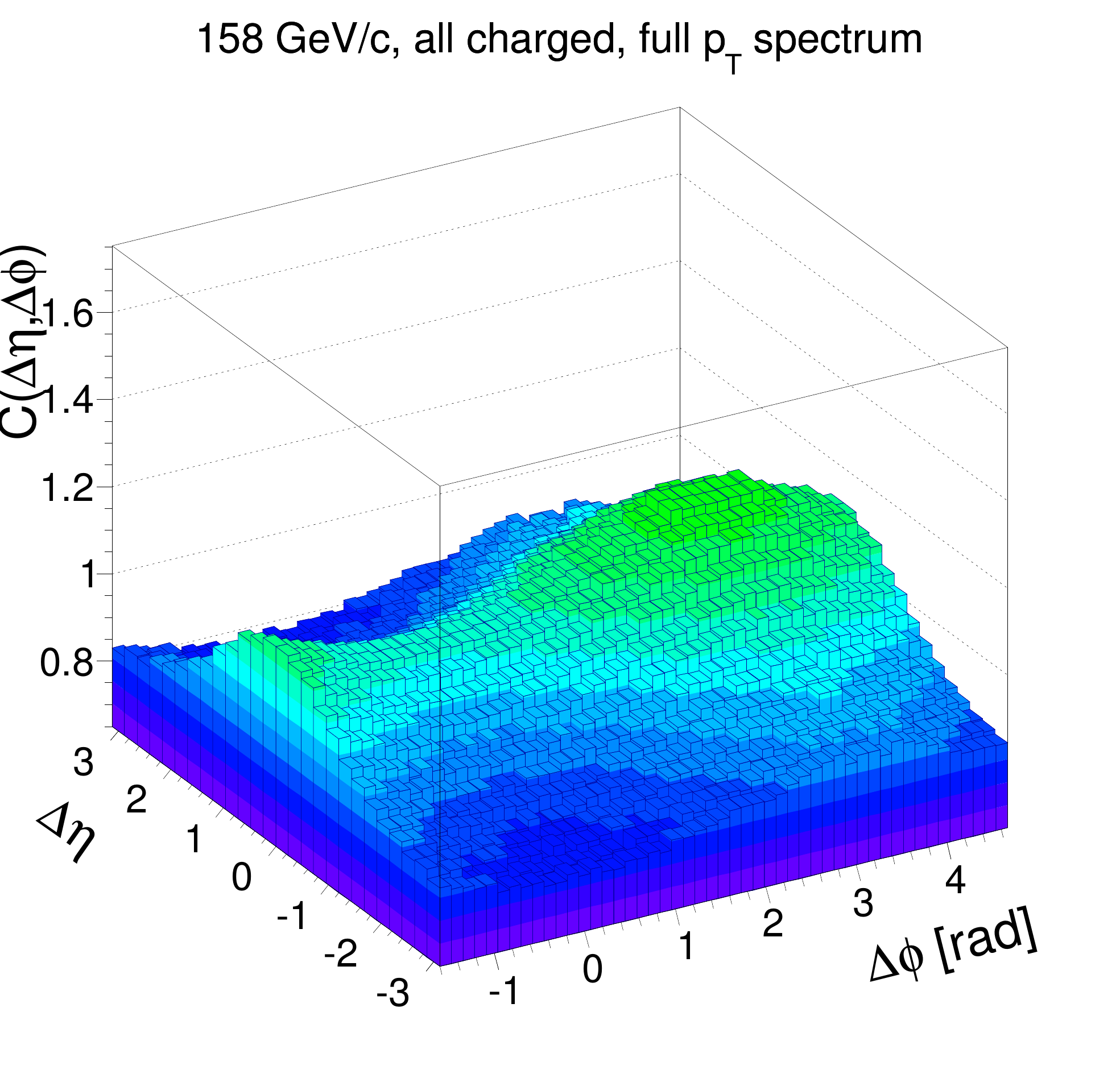}
  \caption{The effect of a transverse momentum cut $p_T < 1.5$~\GeVc on the measured correlation function.}
  \label{fig:ptcut_effect_data}
\end{figure}

\begin{figure}
  \centering
  \includegraphics[width=0.45\textwidth]{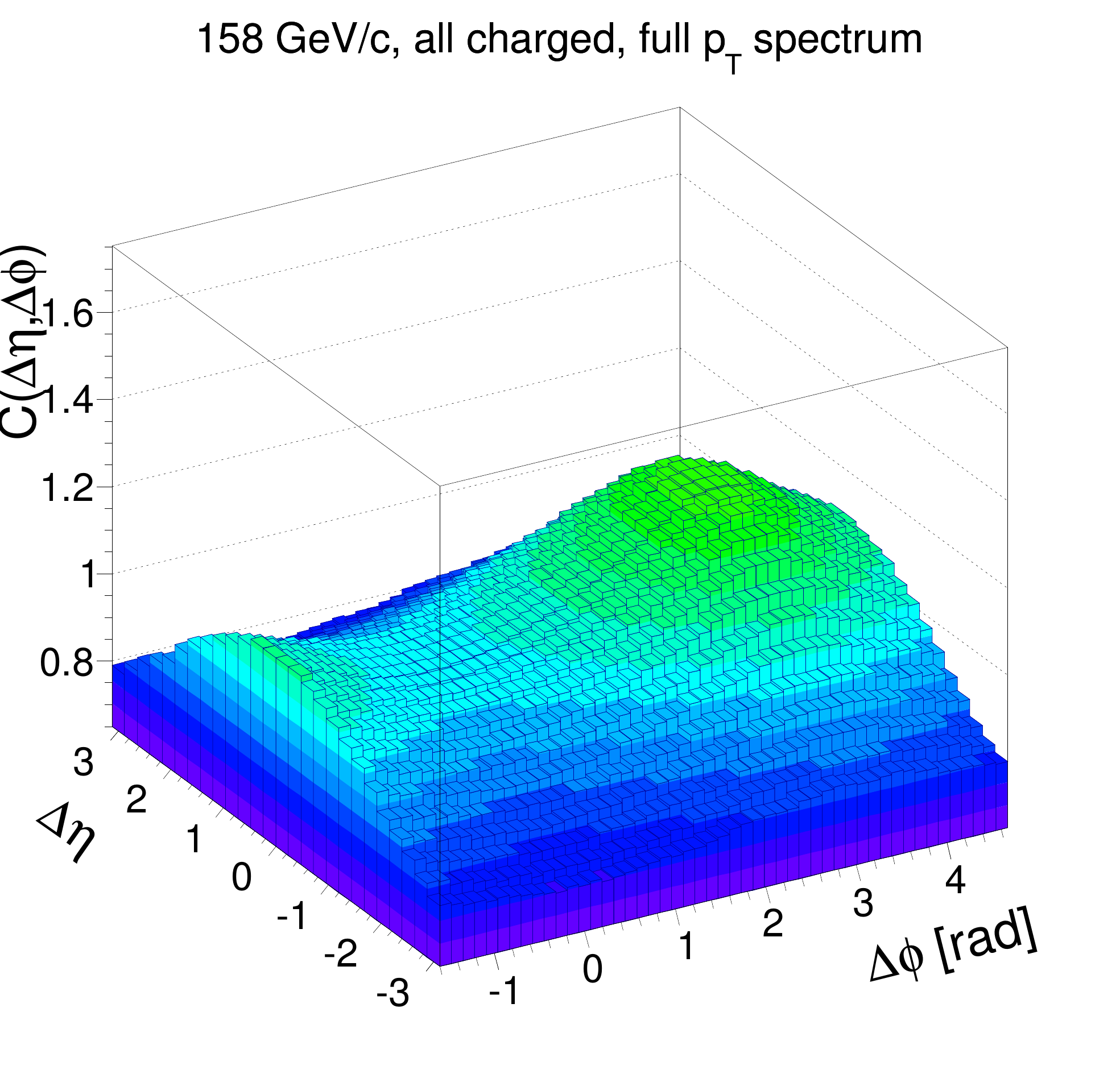}
  \includegraphics[width=0.45\textwidth]{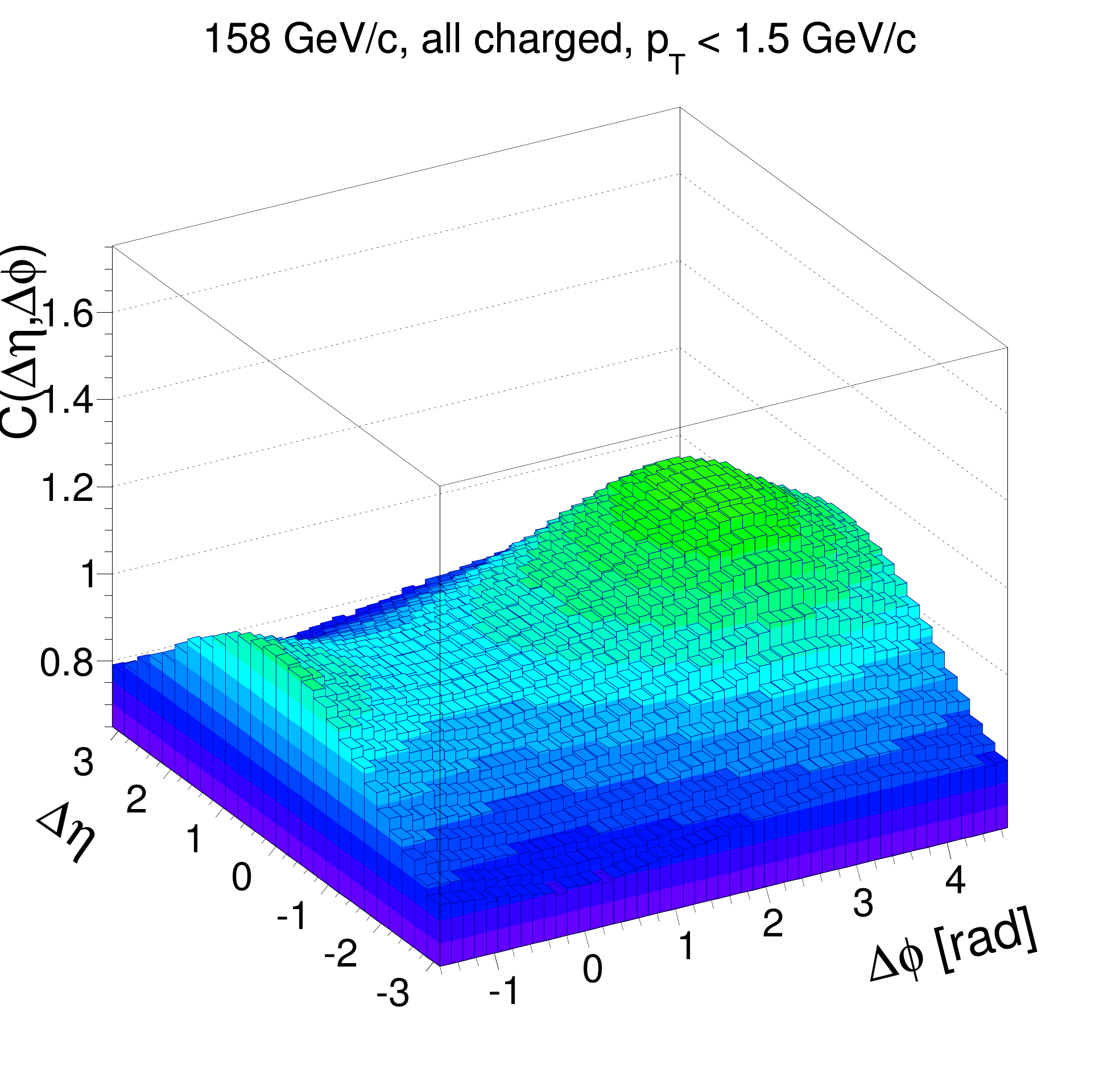}
  \caption{The effect of a transverse momentum cut $p_T < 1.5$~\GeVc on the correlation function obtained from the \Epos model.}
  \label{fig:ptcut_effect_epos}
\end{figure}

Originally, \NASixtyOne meant to study the effects of conservation laws, resonance decays, etc. on the correlation function $\Delta\eta\Delta\phi$. The effects of jets, though expected to be negligible, were reduced by the transverse momentum cut $p_T < 1.5$~\GeVc. An additional analysis was performed to check the effect of this cut. A comparison of the left and right plots in Figs. \ref{fig:ptcut_effect_data} and \ref{fig:ptcut_effect_epos} demonstrates that the influence of the upper $p_T$ cut on the correlation functions for data and the \Epos model, respectively, are indeed negligible. 

\subsection{Comparison with the ALICE experiment}

\begin{figure}[h]
  \centering
  \includegraphics[width=\textwidth]{pics/pp/Data_corrected/Data_corrected_all.pdf}
  \\
  \vspace{1cm}
  \includegraphics[width=\textwidth]{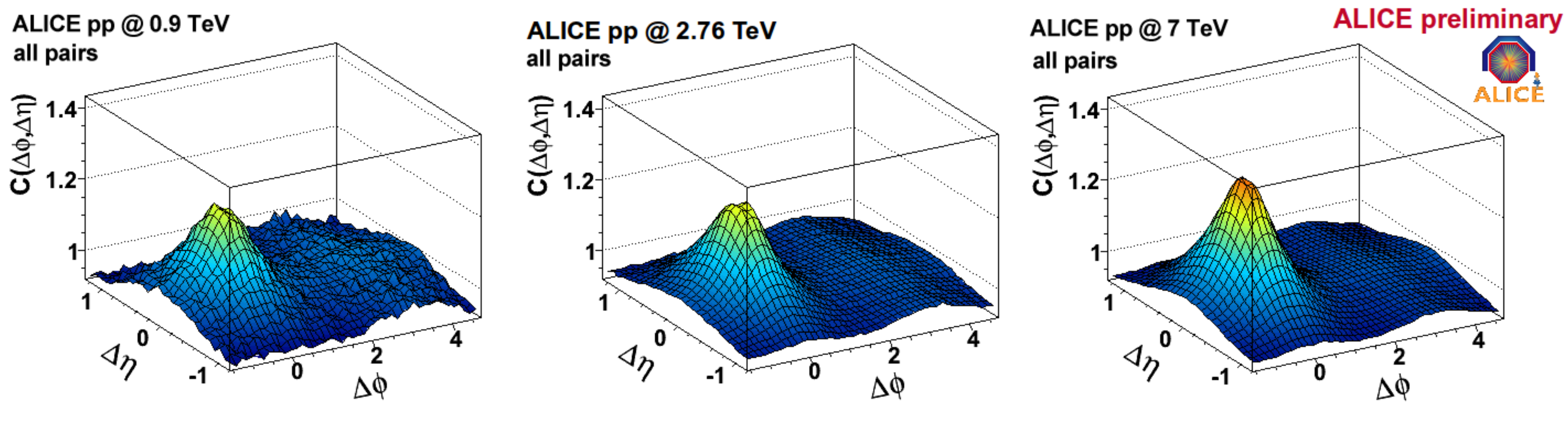}
  \caption{Correlation functions for all charged particle pairs from \NASixtyOne (five upper plots, same as in Fig.~\protect\ref{fig:data_corr_all}) compared with the ALICE results (three lower plots) taken from Ref.~\cite{Janik:2012ya}.}
  \label{fig:na61_vs_ALICE_energydependence}
\end{figure}

The \NASixtyOne  results are compared to the results from the ALICE experiment in Fig.~\ref{fig:na61_vs_ALICE_energydependence}. \NASixtyOne results show a stronger enhancement at $\Delta\phi \approx \pi$ and a significantly weaker maximum at $\Delta\phi \approx 0$ (``jet peak'').

\begin{figure}
  \centering
  \includegraphics[width=0.45\textwidth]{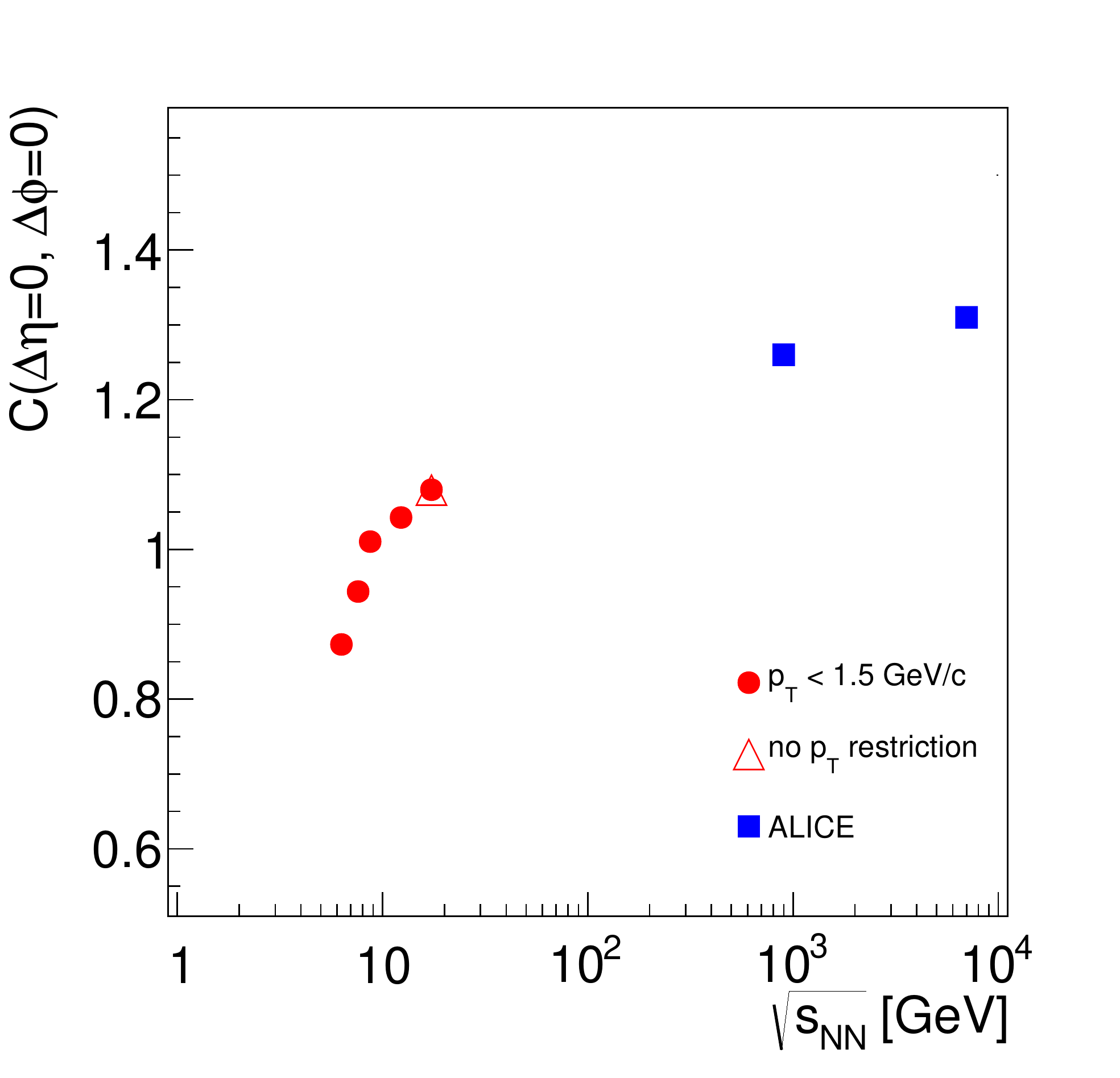}
  \includegraphics[width=0.45\textwidth]{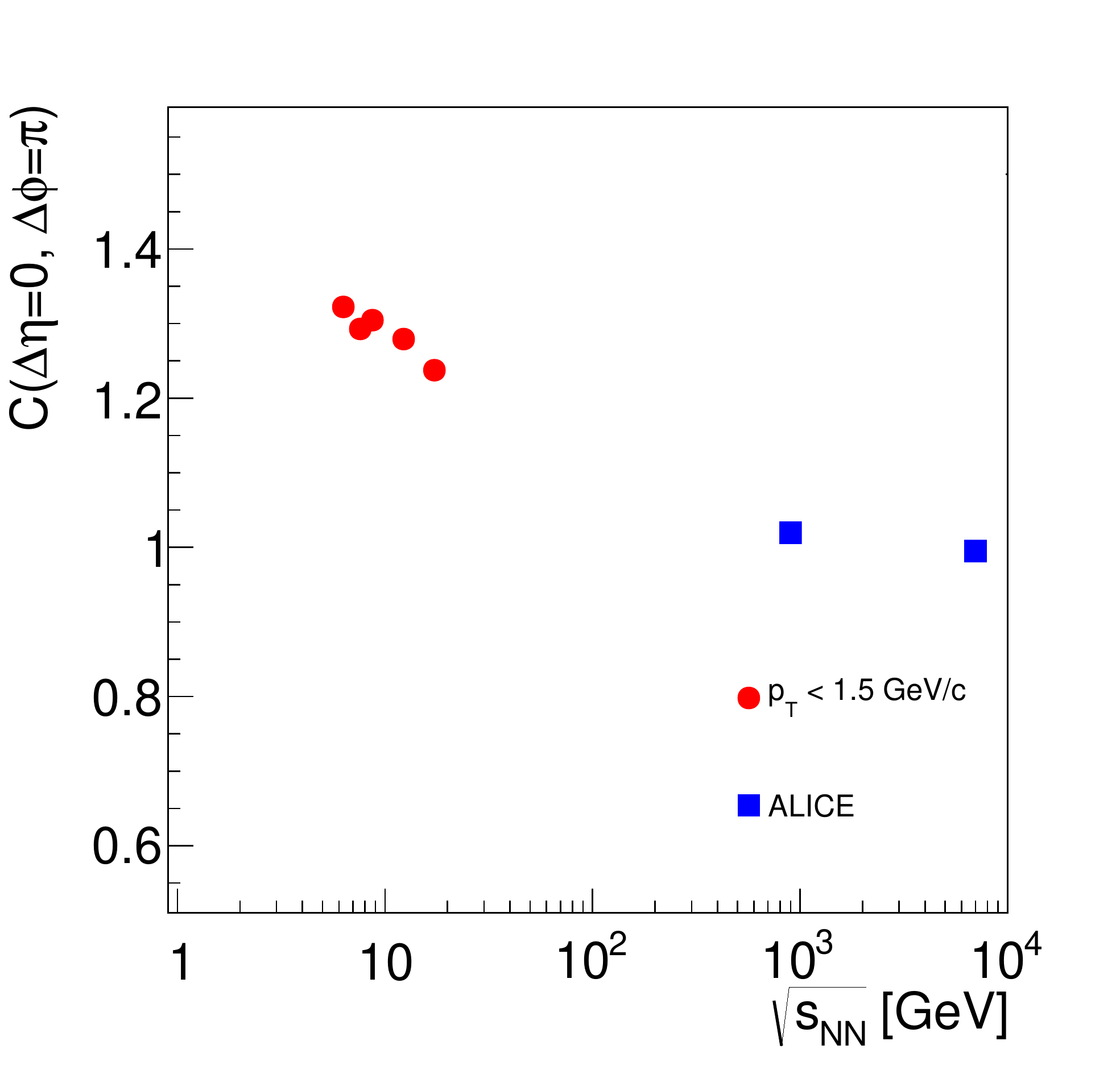}
  \caption{Correlation value $C(\Delta\eta,\Delta\phi)$ for ``jet peak'' and ``resonance hill'' as a function of energy in CMS frame. Left plot shows height of $C(\Delta\eta=0,\Delta\phi=0)$ (``jet peak''). Right plot is a height of $C(\Delta\eta=0,\Delta\phi=\pi)$ (``resonance hill'').}
  \label{fig:c00_c0pi}
\end{figure}

Figure \ref{fig:c00_c0pi} presents the energy dependence of $C(0,0)$ (``jet peak'') and $C(0,\pi)$ (``resonance hill'') from the \NASixtyOne and ALICE experiments. Monotonical behaviour is observed for both $C(0,0)$ and $C(0,\pi)$. $C(0,0)$ rises with collision energy suggesting that the contribution from jets gets stronger with energy. On the other hand, $C(0,\pi)$ decreases with collision energy. This can be explained by a decreasing contribution coming from resonance decays.

\noindent
{\bf Acknowledgements:} This work was supported by the the National Science Centre, Poland grant 2012/04/M/ST2/00816.

\end{document}